\documentclass[%
 reprint, 
 superscriptaddress,
 amsmath,amssymb,
 aps,
]{revtex4-2}    

\usepackage[authormarkup=superscript, defaultcolor=red,hypertexnames=false]{changes} 

\usepackage{graphicx}
\usepackage{dcolumn}
\usepackage{bm}
\usepackage{fixltx2e} 
\usepackage{siunitx} 
\usepackage{tabularx}   
\usepackage{booktabs}   
\usepackage{array}      
\usepackage{ltxcmds}   

\usepackage{dcolumn}

\usepackage{array}      

\usepackage{xr} 
\usepackage{xcolor}   

\newcommand{\mycomment}[1]{}   

\usepackage[colorlinks=true,linkcolor=blue,allcolors=blue]{hyperref}  

\usepackage{lineno}     
 
\newcommand{\lchev}{\left\langle} 
\newcommand{\rchev}{\right\rangle}

\newcommand{\pin}{ p_{\textrm{in}} }  
\newcommand{\durec}{ \delta u_{\textrm{rec}} }    
\newcommand{\durecCrit}{ \delta u_{\textrm{rec}}^{\textrm{crit}} }    

\newcommand{\fr}{\langle \sigma_{i} \rangle}  
\newcommand{\frj}{\langle \sigma_{j} \rangle}  
   
\newcommand{\aveTwo}{\langle \sigma_{i}\sigma_{j} \rangle}    
\newcommand{\fcAve}{\langle f_m \rangle}   
\newcommand{\fcAveData}{\langle f_m \rangle_{\textrm{data}}}    
\newcommand{\fcAveMEM}{\langle f_m \rangle_{\textrm{MEM}}}

\newcommand{\Pdata}{ P_{\textrm{data}}( \bsigma ) }      
          
\newcommand{\PMEM}{ P_{\textrm{MEM}}( \bsigma ) }    
                
\newcommand{\bsigma}{ \bm{\sigma} }

\newcommand{\Dkl}{ D_{\textrm{KL}} }

\newcommand{\aveSigmaData}{\langle \sigma_{i} \rangle_{\textrm{data}} }    
\newcommand{\aveSigmaMEM}{\langle \sigma_{i} \rangle_{\textrm{MEM}} }  
\newcommand{\CijData}{ C_{ij} ^{\textrm{data}} }    
      
\newcommand{\TijkData}{ T_{ijk} ^{\textrm{data}} }    
\newcommand{\TijkMEM}{ T_{ijk} ^{\textrm{MEM}} }     
\newcommand{\PKData}{ P(K)_{\textrm{data}}  }  
\newcommand{\PKMEM}{ P(K)_{\textrm{MEM}}  }   
\newcommand{\aveTwoData}{\langle \sigma_{i}\sigma_{j} \rangle_{\textrm{data}}  }  
\newcommand{\aveTwoMEM}{\langle \sigma_{i}\sigma_{j} \rangle_{\textrm{MEM}}  }

\newcommand{\PK}{ P(K) } 
     
\newcommand{\maxCv}{ \textrm{max}[ C_{v} ] }    
   
\newcommand{\maxChi}{ \textrm{max}[ \chi ] }   
    
\newcommand{\Tmax}{ T_{ \textrm{max} } }    
\newcommand{ \Nc }{ N_c }    
\newcommand{ \BMLearningRate }{ \theta }

\begin{document}

\title{ Maximum entropy models of neuronal populations at and off criticality }
 
\author{T. S. A. N. Sim\~oes}
\affiliation{University of Campania “Luigi Vanvitelli”, Department of Mathematics and Physics, Caserta, Viale Lincoln, 5, 81100, Italy}

\author{F. Lombardi} 
\affiliation{Department of Biomedical Sciences, University of Padova, Padova 35131, Italy}

\author{D. Plenz} 
\affiliation{Section on Critical Brain Dynamics, National Institute of Mental Health, Porter Neuroscience Research Center, Rm 3A-1000, 35 Convent Drive, Bethesda, MD, 20892, USA} 

\author{H. J. Herrmann}
\affiliation{Universidade Federal do Cear\'a, Departamento de F\'isica, Fortaleza, Cear\'a, 60451-970, Brazil}%
\affiliation{ESPCI, PMMH, Paris, 7 quai St. Bernard, 75005, France}   

\author{L. de Arcangelis}
\affiliation{University of Campania “Luigi Vanvitelli”, Department of Mathematics and Physics, Caserta, Viale Lincoln, 5, 81100, Italy}%

\begin{abstract}

Empirical evidence of scaling behaviors in neuronal avalanches suggests that neuronal populations in the brain operate near criticality. Departure from scaling in neuronal avalanches has been used as a measure of distance to criticality and linked to brain disorders. 
A distinct line of evidence for brain criticality has come from thermodynamic signatures in maximum entropy (ME) models.
Both of these approaches have been widely applied to the analysis of neuronal data. 
However, the relationship between  deviations from avalanche criticality and  thermodynamics of ME models of neuronal populations remains poorly understood. To address this question, we study spontaneous activity of organotypic rat cortex slice cultures  in physiological  and drug-induced hypo- or hyper-excitable conditions, which are classified as critical, subcritical and supercritical based on avalanche dynamics. We find that static ME models inferred from critical cultures show signatures of criticality in thermodynamic quantities, e.g.\ specific heat. However, such signatures are also present and equally strong in models inferred from supercritical cultures---despite their altered dynamics and poor functional performance. On the contrary, ME models inferred from subcritical cultures do not show thermodynamic hints of criticality. Importantly, we confirm these results using an interpretable neural network model that can be tuned to and away from avalanche criticality. Our findings indicate that static maximum entropy models, although not constraining dynamical features, correctly distinguish  subcritical from  critical/supercritical systems. However, they may not be able to discriminate between avalanche criticality and supercriticality, suggesting that dynamics is relevant to capture the supercritical behavior and distinguish it from criticality.

\end{abstract} 

\maketitle

\section{\label{sec:Introduction} Introduction }    

Biological neural networks need to perform complex functions and continuously adapt. Such abilities rely on cooperative effects among local and distributed neuronal populations, which underlie the emergence of a variety of collective behaviors in the brain. The analogy between collective behaviors in populations of neurons and cooperative phenomena in physical systems undergoing a phase transition suggests that brain networks self-organize to operate at or near criticality \cite{cragg_organisation_1954,de_Arcangelis_2014}, a state that  provides several functional advantages \cite{kinouchi_optimal_2006,shew_information_2011}.    
This hypothesis is primarily supported by observations of long-range spatio-temporal correlations and neuronal avalanches across species and spatial scales.
Neuronal avalanches are cascades of local synchronized activity whose size and duration distributions follow power law behaviors,  hallmarks of criticality that imply  absence of characteristic scales. First identified in acute slices and long term slice cultures of rat cortex \textit{in vitro} \cite{beggs_neuronal_2003},  
neuronal avalanches have since been reported \textit{in vivo} in rats \cite{gireesh_neuronal_2008}, monkeys \cite{petermann_spontaneous_2009}, and  other species \cite{hahn_spontaneous_2017,ponce-alvarez_whole-brain_2018}, including MEG and EEG recordings of the human brain \cite{shriki_neuronal_2013,lombardi_long-range_2021,scarpetta_criticality_2023}.
Concomitant evidence of tuning to criticality in neuronal systems has come from maximum entropy (ME) models of neural activity, which focused on thermodynamic aspects such as the divergence of the specific heat  \cite{tkacik_thermodynamics_2015,mora_dynamical_2015,hahn_spontaneous_2017,lotfi_signatures_2020,sampaio_filho_ising-like_2024}.

ME modeling has proven a powerful approach to study  the dynamics of biological neural networks \cite{schneidman_weak_2006,tkacik_spin_2009,tkacik_searching_2014}.  
The spiking activity of neurons can be described as a binary process, $\sigma (t)$, in which $\sigma = \pm1$ represents the state of the neuron at a given time $t$, i.e.\ spiking  for $\sigma = 1$ and silent for $\sigma = -1$ \cite{rieke_spikes_1999}. This approach allows one to define an empirical distribution of binary activity patterns, $P(\bf{\sigma})$, which specifies the probability of observing a given spiking pattern in a population of neurons. For $N$  neurons, the distribution $P(\bf{\sigma})$ of all possible $2^N$  spiking patterns (or states) fully characterizes  population dynamics. However, estimating $P(\bf{\sigma})$ directly from data is often impractical since the number of possible  states grows exponentially with the number of neurons \cite{mora_are_2011}. Maximum entropy modeling  offers a possible solution to this problem by constructing the least-biased probability distribution that matches a number of selected statistics, such as individual firing rates and pairwise correlations, while maximizing the entropy of $P(\bf{\sigma})$. The resulting distribution is mathematically equivalent to the Boltzmann distribution \cite{schneidman_weak_2006,tkacik_spin_2009,gardella_modeling_2019}, which can then be analyzed using  thermodynamic tools, such as the fluctuation-dissipation theorem \cite{tkacik_thermodynamics_2015,sampaio_filho_ising-like_2024,simoes_thermodynamic_2024}. 

This procedure can be interpreted as building a thermodynamics-like framework to describe steady-state properties (e.g., average firing rates in neural networks) of non-equilibrium systems \cite{hansen_statistical_2023, hansen_thermodynamics-like_2025}.  ME modeling has been  extensively applied to neuronal data, from the evoked activity in the salamander retina \cite{schneidman_weak_2006,tkacik_searching_2014,tkacik_thermodynamics_2015,humplik_probabilistic_2017} and the nervous system of the C.\ Elegans \cite{chen_searching_2019}, to  \textit{in vivo} and \textit{in vitro}  populations of rodent cortical neurons  \cite{sampaio_filho_ising-like_2024}. In this context, criticality is identified by a maximum in the susceptibility or in the specific heat of the data-inferred model close to unit temperature, corresponding to the observed statistics of neural activity. Such signatures of criticality in neural data are typical of equilibrium phase transitions and can be considered static,  as opposed to those coming from scaling of neuronal avalanches, which are intrinsically dynamic and associated to non-equilibrium phase transitions \cite{Tauber_2014,henkel2008non}.
Recently, ME models that take into account temporal dynamics of neural activity have been proposed to enhance (static) evidence of criticality, while matching avalanche statistics  \cite{mora_dynamical_2015}.

However, to what extent static and dynamic signatures of criticality in neural systems need to coexist and agree in baseline normal condition and how they change  when such systems are perturbed away from baseline condition remains poorly understood. How  do avalanche scaling signatures translate into features of static maximum entropy models? Are deviations from dynamical criticality captured by changes in thermodynamic properties of the corresponding maximum entropy models and vice versa? As criticality is increasingly used as a biomarker for brain and behavioral disorders \cite{Zimmern_2020,fl2020jneurosci,Huo_LC_2024,shew25neuron}, these questions also become  of key relevance  in neuroscience. 

Here, we address them by analyzing neural cultures in baseline and pharmacologically perturbed conditions, together with an interpretable neural network model whose dynamics can be tuned to  criticality \cite{michiels_van_kessenich_critical_2018,simoes_thermodynamic_2024}.
We  consider three sets of neuronal cultures: one in baseline,  physiological condition, one with reduced neural excitability, and one with reduced neural inhibition \cite{shew_neuronal_2009}. 
Cultures in physiological condition showed neuronal avalanches whose size and duration distributions were consistent with power-laws  and were classified as critical in \cite{shew_neuronal_2009}. In cultures  with reduced excitability instead, avalanches were small and short-lived, with  exponential size and duration distributions. These cultures  were identified as subcritical \cite{shew_neuronal_2009}. In contrast, disinhibited cultures exhibited high excitability and  a pronounced increase in the probability of very large avalanches---of the order of the system size---, leading to bimodal-like size/duration distributions. These cultures were classified as supercritical \cite{shew_neuronal_2009}. Importantly, each of these scenarios can be reproduced with our network model by tuning a single parameter.

We use a maximum entropy approach to model experimental and numerical data by constraining firing rates,  pairwise correlations, and the distribution of synchrony, $P(K)$, defined as the probability that $K$ neurons are active simultaneously \cite{tkacik_searching_2014}. By studying their thermodynamic properties, we show that the inferred models (mostly) agree with the dynamical classification of hypoexcitable cultures as subcritical, and correctly distinguish them from dynamically critical/supercritical cultures. However, they are not able to discriminate between dynamical criticality and supercriticality. 
These results are confirmed by the analysis of our integrate-and-fire (IF) neural network, whose inferred ME model closely matches the one inferred from neuronal cultures, despite  the  very simplified structure of the IF neural network.

The paper is organized as follows. In Sec.\ \ref{sec:Data}, the neural network  model, the experimental setup and data, and the quantities considered for the ME modeling are described. In Sec.\ \ref{sec:Ising}, the ME modeling method is explained and the inferred ME distributions are analyzed for both the numerical and experimental datasets. Finally, Sec.\ \ref{sec:Discussion} summarizes and discusses the results.

\section{\label{sec:Data} Data acquisition and Methods}

\subsection{\label{subsec:IFModel} Integrate-and-fire neuronal network model} 

\paragraph*{Network dynamics.}We consider an integrate-and-fire (IF) model    with synaptic plasticity and  a refractory period of one timestep, during which neurons remain inactive after firing \cite{michiels_van_kessenich_critical_2018}.  We implement the model on scale-free, directed networks with different numbers of neurons, $N$, placed within a cube of side $L$, but keep the density $N/L^3=0.016$ constant \cite{nandi_scaling_2022}. The behavior of the model, such as neuronal avalanche dynamics,   in practice does not  depend on the topology of the network \cite{lombardi_balance_2017}. A fraction $\pin = 20 \%$ of neurons is inhibitory \cite{zeraati_self-organization_2021}. The out-degree $k$ of each neuron follows a power-law distribution, $P(k) \propto k^{-2}$, with $k \in [2,20]$ for systems with $N \leq 100$ and $k \in [2,100]$ otherwise. The probability that two neurons are connected decays exponentially with the Euclidean distance $r$, $P(r) \propto e^{-r/r_{0}}$,  where $r_0 = 5$ \cite{roerig_relationships_2002}. The resting potential of each neuron $i$ is set at $v_{i} = 0$. A neuron fires when $v_{i} \geq v_{c} = 1$, transmitting signals to all its post-synaptic neurons $j$ according to the following equation: 
\begin{align}	    
\label{eq:synaptic-transmition} 	v_j (t+1) &= v_j(t) \pm v_i(t) u_i(t) g_{ij}  \textrm{ ,} \\    
\label{eq:STP}	u_i (t+1) &= u_i(t) \cdot (1 - \delta u) \textrm{ ,} \\    
\label{eq:v-reset} 	v_i (t+1) &= 0 \textrm{ ,}    
\end{align}       
where $+$ and $-$ are for excitatory and inhibitory pre-synaptic neurons,  $g_{ij}$ is the strength of the synapse connecting $i$ to $j$, and $u_i$ indicates the synaptic resources of the pre-synaptic neuron $i$. The constant $\delta u = 0.05$ controls the fraction of  neurotransmitters released \cite{ikeda_counting_2009}.

We start our simulations  with the synaptic strengths uniformly distributed in the interval $g_{ij} \in [0.4,0.6]$ and with $u_{i} = 1$ for all neurons. To sustain network activity, a small external input $\delta v=0.1$ is added to a random neuron at each time step \cite{das_critical_2019}. A timestep corresponds to the time interval between the generation of the action potential in the pre-synaptic neuron and the change in the membrane potential of the postsynaptic one, and is of the order of 10 ms  \cite{benayoun_avalanches_2010,boudkkazi_release-dependent_2007,lisman_sequence_2007}. When a neuron reaches the threshold $v_c$, it fires an action potential and  propagates activity to other neurons, causing an avalanche to start. An avalanche ends as soon as $v<v_c$ for all neurons.  Triggering of subsequent avalanches is ensured by the small external input $\delta v$.

After each avalanche, the pool of neurotransmitters of each neuron, $u_i$, is replenished by an amount $\durec$, i.e.\ $u_{i}(t) \rightarrow u_{i}(t) + \durec $. Recovery is implemented between avalanches according to the separation of time scales in self-organized criticality  models, in which  avalanches are considered as  almost instantaneous events. For a given $N$, the system can be set to criticality by tuning $\durec$ to a certain value $\durecCrit (N)$. At criticality, the system exhibits avalanches with size $S$ and duration $D$ distributed according to power-laws whose cut-off scales with the system size $N$. Setting $\durec < \durecCrit (N)$  leads to subcritical dynamics, characterized by an exponential decay in the distributions of avalanche sizes, $P(S)$, and avalanche durations, $P(D)$. Conversely, $\durec > \durecCrit (N)$ sets the system  in a super-critical state where there is a sharp increase in large and long avalanches, leading to the appearance  of local maxima in $P(D)$ and in $P(S)$ around the power-law cut-off (SI, Fig.\ \ref{figS:aval-dists} for the distributions and Table \ref{tableS:durec} for the set of values $\durecCrit$). 
For simulations of subcritical and supercritical IF networks, we set $\durec(N) = 0.1 \cdot \durecCrit(N)$ and $\durec(N) = 10 \cdot  \durecCrit(N)$, respectively. 

\paragraph*{Plasticity rule and distribution of synaptic strength.}   
Before performing measurements, we apply the following synaptic plasticity rule for $10^4$ avalanches or until  one  $g_{ij}$ first reaches the value $g_{\textrm{min}} =10^{-5}$: we increase the strength of the synapses $g_{ij}$ proportionally to the voltage variation induced in the post-synaptic neuron $j$ due to $i$ as $g_{ij} (t + 1) = g_{ij} (t) + \delta g_{j}$, where $ \delta g_{j} = \beta \lvert v_{j} (t + 1) - v_{j} (t) \rvert $, with $ \beta = 0.04 $ setting the plasticity rate;  then, at the end of each avalanche, we decrease all $g_{ij}$ by the average increase in strength per synapse, $ g_{ij}(t+1) = g_{ij}(t) - \frac{1}{N_s} \sum{ \delta g_{j} } $, where $N_{S}$ is the number of synapses.    Synapses that are rarely active tend to weaken over time \cite{bi_synaptic_1998}. This plasticity routine mimics long-term adaptive processes  that occur on timescales much longer than usual activity recordings. For this reason, we apply it only before recording network activity to shape  the distribution of synaptic strengths, which is then kept fixed, and refer to it as long-term adaptation as opposed to the short-term changes in synaptic resources.

\paragraph*{Simulations.} For each system size $N$ and state of the dynamics (subcritical, critical, and supercritical), we generate five independent  time series of neuronal spikes, each corresponding to a different realization of the network configuration. We refer the reader to the SI for a comprehensive account of  the avalanche size and duration distributions in each network state (SI, Fig.\ \ref{figS:aval-dists}). Distributions of avalanche size and duration are independent of the external drive, $\delta v$, over a wide range of values (SI, Fig.~\ref{figS:aval-dists-vs-dv-and-nstim}a, b), and weakly depend on the number of simultaneously stimulated neurons, $n_{stim}$---as far as $n_{stim} \lesssim 20\%$ of $N$ (SI, Fig.~\ref{figS:aval-dists-vs-dv-and-nstim}c, d).

\subsection{\label{subsec:ExpData} Experimental methods and data}      

The data analyzed here were selected from a set of recordings for a previously published study \cite{shew_neuronal_2009}, in which further details can be found. Organotypic coronal slices of rat somatosensory cortex (350 µm thick, postnatal day 0–2; Sprague Dawley), co-cultured with midbrain tissue (ventral tegmental area; 500 µm thick), were maintained on a planar $8 \times 8$ microelectrode array (MEA) as described in \cite{shew_neuronal_2009}. Of the 64 microelectrodes in the array, the four corner electrodes were excluded from the analysis, leaving $N=60$ active channels for actual data collection. 
To induce a supercritical state, the GABA\textsubscript{A} receptor antagonist picrotoxin (PTX) was added to the culture medium. 
The subcritical state was induced instead by applying two types of glutamate receptor antagonists that reduced  excitatory transmission. Cultures were treated either with (2R)-amino-5-phosphonovaleric acid (AP5), an NMDA receptor antagonist, or with a combination of AP5 and 6,7-dinitroquinoxaline-2,3-dione (DNQX), an AMPA receptor antagonist. Cultures without any drug treatment (baseline) were identified as being in a critical state. 
Using a recording head stage inside the incubator (MEA1060 w/blanking circuit; $ \times $1200 gain; bandwidth 1–3000 Hz; 12 bit A/D; range 0–4096 mV; Multi Channel Systems), the local-field potential (LFPs; 4 kHz sampling rate; reference electrode in bath) was obtained from 1 hour recordings of extracellular activity (low-pass, 100 Hz, phase-neutral) of the same culture. For each electrode, negative peaks in the LFP exceeding four standard deviations of the electrode noise were identified as firing events, and their timing was recorded. In this work,  15 recordings were analyzed, 5 for each condition: critical, subcritical (1 with AP5, 4 with mixed AP5/DNQX), and supercritical. Distributions of avalanche sizes extracted from the whole set of recordings are reproduced from \cite{shew_neuronal_2009} and shown in Fig.\ \ref{figS:aval-dists-exp} of the SI.

\subsection{\label{subsec:firingStatistics} Quantification of neural dynamics}   
  
For the experimental datasets, each of the approximately one hour long recordings is divided into time bins of duration $\Delta t_b = \SI{25}{ms}$. This is  within the range of values used in previous ME analysis of  neuronal data \cite{schneidman_weak_2006,tkacik_searching_2014,tkacik_thermodynamics_2015,mora_dynamical_2015,meshulam_successes_2023}, and gives a total of $N_{b} \approx 1.4 \cdot 10^{5} $ time bins for each recording. For the numerical datasets, we have $N_{b} = 10^{7}$ time bins, each of duration $\Delta t_b = 5$ time steps. We consider the numerical timestep to be on  the order of a few milliseconds. Setting  $\Delta t_b = 5$ provides a  time bin for numerical data that is approximately equivalent to that used for  experimental data.  
For each electrode on the MEA and neuron in the model, a binary variable $\sigma_{i}(k)\in\{-1,1\}$ is assigned to each time bin $k$, with $\sigma_{i}(k)=1$ if the neuron or electrode $i$ fires at least once in the bin $k$ and $\sigma_{i}(k)=-1$ otherwise. Thus, we do not count multiple spikes that may occur in a time bin. We note that, for the selected $\Delta t_b$, the average number of spikes per neuron in non-silent time bins, $\langle n_s | n_s > 0 \rangle$,  is approximately one in all the three states of both experimental and numerical networks (SI, Fig.\ \ref{figS:spike-per-bin-dists}).  In the critical and subcritical state,  $\langle n_s | n_s > 0 \rangle$ stays nearly constant in a wide range of $\Delta t_b$ (SI, Figs.\ \ref{figS:spike-per-bin-dists}a, b, d, e). Remarkably, numerical and empirical values are very close. In the supercritical state, $\langle n_s | n_s > 0 \rangle$ stays close to unity around  $\Delta t_b = 25$ ms, and reaches $\approx 1.6$ for $\Delta t_b = 100$ ms (SI, Fig.\ \ref{figS:spike-per-bin-dists}f). These properties are robust with respect to  external drive and number of stimulated neurons in the IF network model (SI, Fig.\ \ref{figS:spike-per-bin-vs-dv-and-n_stim}).   
 
To characterize neural dynamics, we consider three quantities: the average activity of each neuron or electrode $i$, the two-point correlation function between each pair $(i,j)$, and the probability of synchrony, $P(K)$.
The average activity of a neuron/electrode $i$ is defined as  
\begin{equation} \label{eq:averageLocalActivity} 
    \langle \sigma_i \rangle = \frac{1}{N_{b}} \sum_{k=1}^{N_{b}} \sigma_i ( k ) \textrm{ ,}       
\end{equation} 
and is related to the firing rate $r_i$,  i.e.\ $r_{i} = ( \langle \sigma_i \rangle + 1 ) / 2 \Delta t_{b} $ \cite{tkacik_searching_2014}. The distributions of $\langle \sigma_i\rangle$ show that activity is very sparse in the subcritical state, slightly increases in the critical state, and becomes more intense in the supercritical state (SI, Fig.\ \ref{figS:aveSigma-dists}). For the chosen parameters, the neural network model shows average activity rates close to the empirical ones (SI, Fig.\ \ref{figS:aveSigma-dists}). In a given network state, the average firing rate can be tuned, to some extent, by modulating the external drive or the number of simultaneously stimulated neurons. Importantly, activity sparseness is preserved for a wide range of external drive strengths and stimulated neurons (SI, Figs.\ \ref{figS:dists-vs-dv}, \ref{figS:dists-vs-n_stim}), in particular in the critical and subcritical state, consistently with ongoing activity in populations of neurons.

The average two-point activity correlation for each of the $N \cdot ( N - 1 ) / 2$ distinct pairs of neurons/electrodes, $(i,j)$, is defined as       
\begin{equation} \label{eq:averageTwoPointActivity} 
    \langle \sigma_i \sigma_j \rangle = \frac{1}{N_{b}} \sum_{k=1}^{N_{b}} \sigma_i ( k ) \sigma_j ( k )  \textrm{ .}       
\end{equation}  
Together with $\langle \sigma_i \rangle$ (Eq.~(\ref{eq:averageLocalActivity})), these quantities define the two-point correlation functions  
\begin{equation} \label{eq:Cij}                                
    C_{ij} = \left\langle (\sigma_i - \fr) \cdot (\sigma_j - \frj) \right\rangle = \aveTwo - \fr \frj \textrm{ ,}         
\end{equation}  
which quantify the tendency of $i$ and $j$ to fire simultaneously. We observe that $C_{ij} \approx 0$ in the subcritical state (SI, Figs.\ \ref{figS:Cij-dists}a, d). Pairwise correlations increase in the critical state and become markedly stronger in the supercritical state (SI, Figs.~\ref{figS:Cij-dists}b, c, e, f), for both empirical and numerical data---independently of $\delta v$ and number of simultaneously stimulated neurons. We note that negative pairwise correlations are much weaker than the positive ones, in particular in the supercritical state.

Finally, the  probability $P(K)$ that $K \in [0,N]$ neurons/electrodes fire simultaneously during the same time bin is given by 
\begin{equation} \label{eq:PK} 
    P(K) = \frac{1}{N_b} \sum_{k=1}^{N_b}  \delta_{K,K'(k)} \textrm{ ,} 
\end{equation} 
where $\delta_{K,K'(k)}$ is the Kronecker delta function and $K'(k) = \sum_{i=1}^{N} \left( \sigma_{i}(k) + 1 \right) / 2$ counts the number of neurons/electrodes that fired during the time bin $k$. We note that, since $P(K)$ is a distribution, only $N$ out of the $N+1$ values of $P(K)$ are independent because of the normalization condition $\sum_{K=0}^{N} P(K) = 1$. The constraint on the overall distribution of activity can be interpreted as a global regulatory  mechanism, e.g.\ inhibitory hubs or other  hidden degrees of freedom contributing to collective behavior of the network,   intrinsic or extrinsic \cite{tkacik_searching_2014}.

\begin{figure*}[!ht]   
	\includegraphics[width=0.9\linewidth]{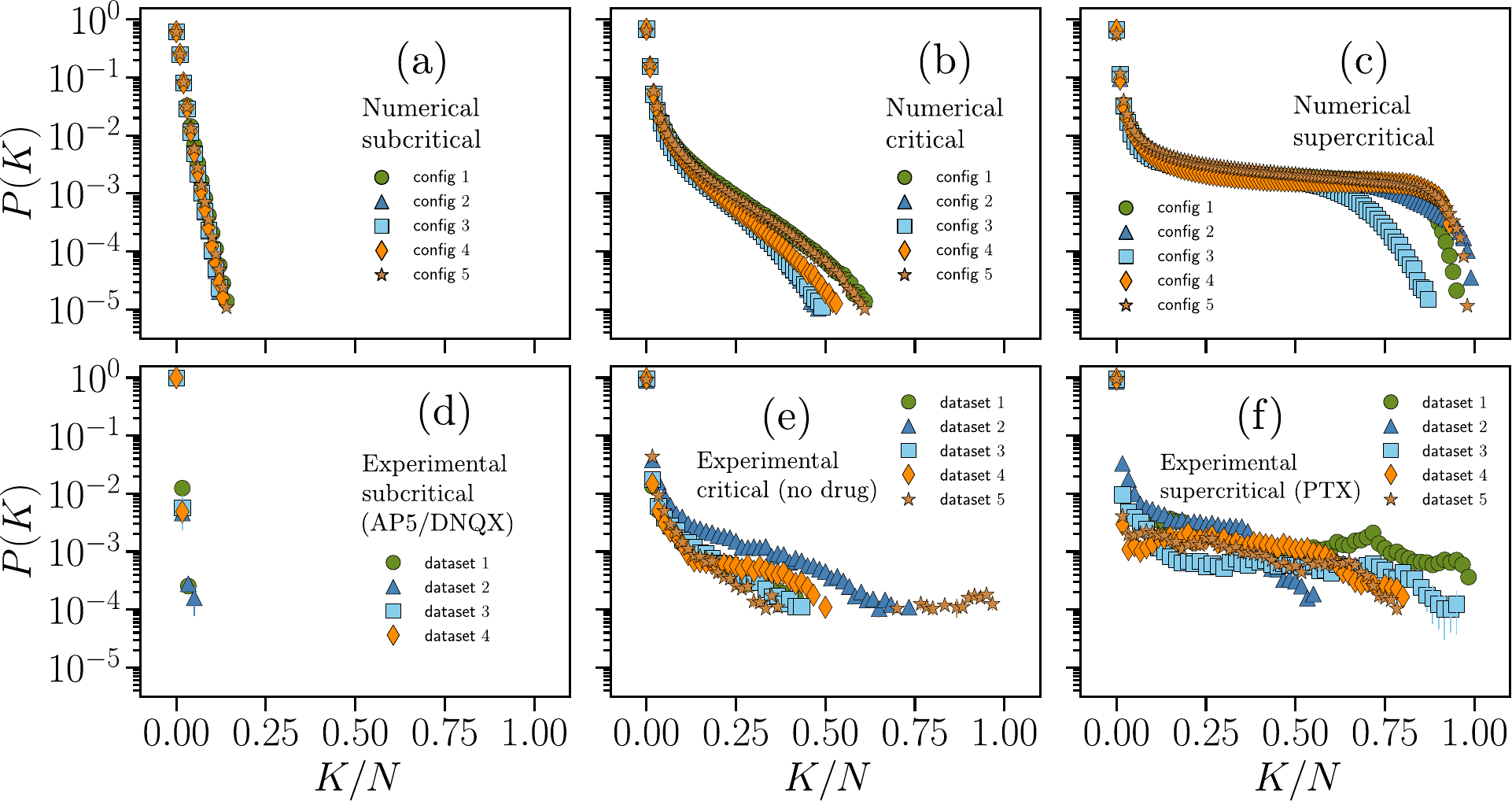}        
	\centering 
	\caption{\textbf{Distributions of synchrony for IF model and cortex slice cultures.} Distribution $P(K)$ for IF networks with $N=100$ neurons and cortex slice cultures in the subcritical (a, d), critical (b, e), and supercritical (c, f) state. 
    Different symbols correspond to either distinct network configurations of the IF model  (a–c) or different  neuronal cultures (d–f). 
    Error bars represent the standard error of the mean estimated from 100 independent subsets of the $N_b$ samples, and are always smaller than or equal to the symbols size.}     
	\label{fig:PK}     
\end{figure*}            

\subsubsection{Distribution of synchrony $P(K)$}\label{subsubsec:PK}

In Fig.\ \ref{fig:PK}, we show the distributions $P(K)$ (Eq.\ \ref{eq:PK}), for both numerical and experimental data in the subcritical, critical, and supercritical regimes. We observe that $P(K)$ exhibits distinctive features depending on the dynamical state. For subcritical networks (Fig.\ \ref{fig:PK}a, d), $P(K)$ decays rapidly, with a maximum $K$ less than 20\% of the total number of neurons. This  indicates   low synchronous  activity in the network,  in line with the exponential decay in the distributions of avalanche sizes and durations  (SI, Fig.\ \ref{figS:aval-dists}d, i). In critical networks instead (Fig.\ \ref{fig:PK}b, e), the decay of $P(K)$ is much slower and non-exponential, indicating that in the critical state  synchronous neuronal activations (spikes) are more likely to occur and  can involve a large fraction of the network ($> 50\%$). The distribution $P(K)$ further broadens in the supercritical state (Fig.\ \ref{fig:PK}c, f), showing  an approximate plateau over a broad range of the fractional synchrony, i.e.\ $ 0.1 \lesssim K/N \lesssim 0.75$,  and decaying rapidly only near the system size $K = N$, i.e.\ $K/N = 1$. This behavior reflects a pronounced increase in  synchronous neural activity that can span the entire network, as also indicated by the sharp maximum at the cut-off in the distributions of avalanche sizes and durations (SI, Figs.\ \ref{figS:aval-dists}b, g). In the neural network model, the average synchrony depends on the number of simultaneously stimulated neurons and, to a lesser extent, on the strength of the external drive, $\delta v$ (SI, Figs.~\ref{figS:dists-vs-dv}c, f and \ref{figS:dists-vs-n_stim}c, f). Importantly,  the maximum fraction of simultaneously active neurons does not increase by increasing these parameters, in particular in the critical state, which is robust and controlled by $\delta u$.

\subsubsection{Allometry of population firing rates}\label{subsubsec:allometry}

\begin{figure}[!ht]        
\centering 
\includegraphics[width=0.8\linewidth]{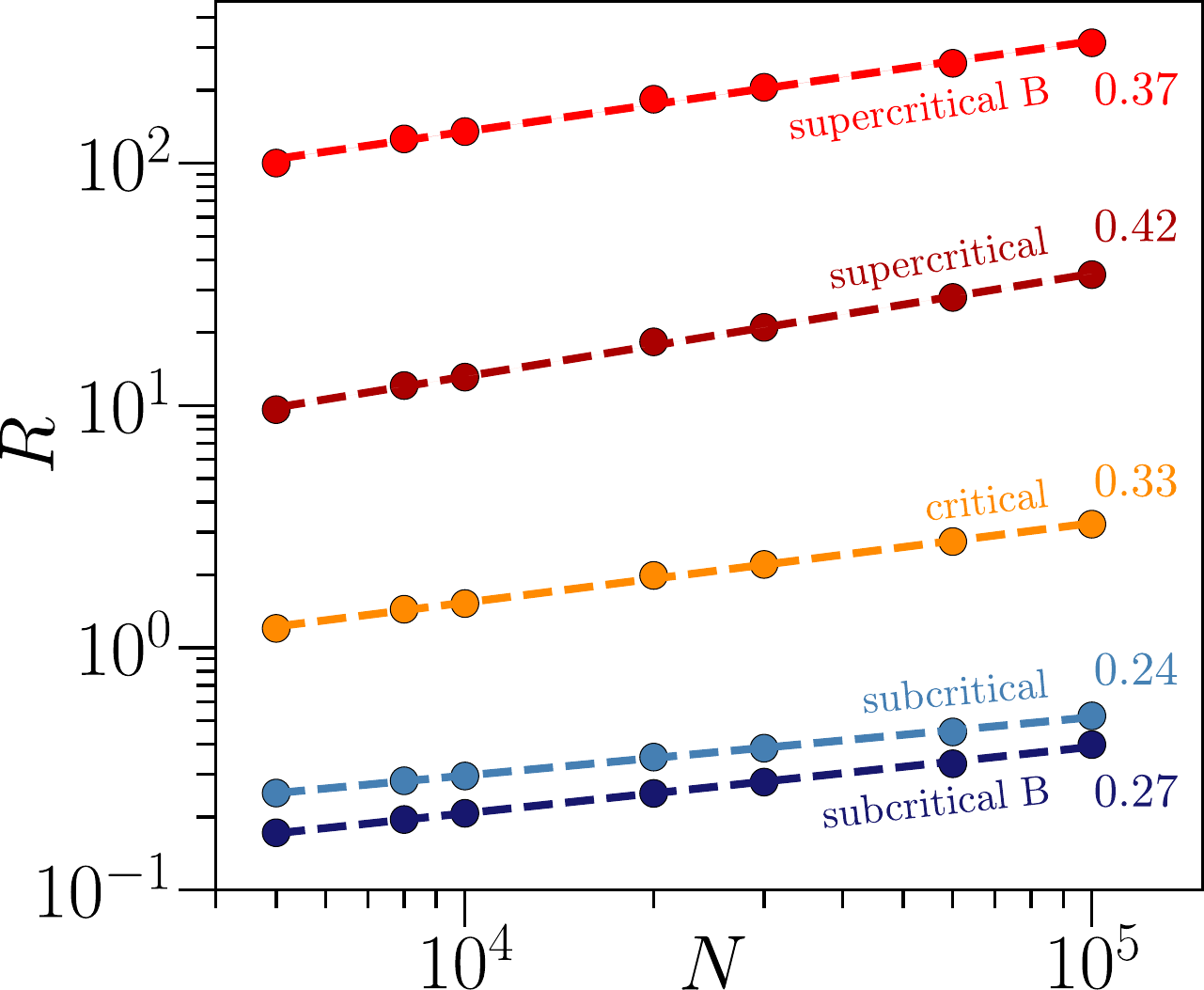} 
	\caption{\textbf{Scaling of the population firing rates with size in the IF network model.} Population firing rate $R$ as a function of the number of neurons, $N$, for sub-critical, critical, and supercritical IF networks. Dashed lines are linear least-square fits of the form $ \log R = \eta \cdot \log N + b$. The $\eta$ value is shown in the figure next to  the corresponding curve. Error is $\approx 0.04$ for all $ \eta $. Subcritical B and supercritical B denote results obtained with $\durec(N) = 0.01 \cdot \durecCrit(N)$ and $\durec(N) = 100 \cdot \durecCrit(N)$, respectively. For each $N$, results are averaged over $2000$ network configurations, and over $ 2 \cdot 10^{4}$ avalanches for each configuration.  
    }
	\label{fig:allometry}     
\end{figure} 

From the firing rates $r_{i} = ( \langle \sigma_i \rangle + 1 ) / 2 \Delta t_{b}$ for individual neurons, one can define the population firing rate $R =\sum_{i=1}^{N}r_{i}$, which in our case corresponds to the average synchrony, $\langle K \rangle$, and measures the number of neurons that are simultaneously active per unit time. Recent studies indicate that this quantity exhibits allometric scaling, a non-trivial scaling relationship with the number of neurons, $R\propto N^{\eta}$ with  $\eta<1$ \cite{simoes_allometric_2026,karbowski_thermodynamic_2009}, which can be derived from the finite-size scaling of neuronal avalanches \cite{simoes_allometric_2026}. 
In our IF model, we observe that $R$ scales sublinearly with $N$  in all the three network states (Fig.\ \ref{fig:allometry}).

The scaling exponent $\eta$ is slightly lower in the subcritical state and higher in the supercritical state compared to criticality  (Fig.\ \ref{fig:allometry}). However, for each fixed network size $N$, the population firing rate  is highest in the supercritical state and lowest in the subcritical state  (Fig.\ \ref{fig:allometry}). 
According to \cite{simoes_allometric_2026}, the robustness of the allometric scaling across network states could be explained by the fact that, in the IF model,  avalanche size and duration distributions 
always have an intermediate power-law regime with similar exponents    and a cutoff that scales similarly with $N$, independently of the tuning parameter $\durec$ (SI, Fig.\ \ref{figS:avalanche-dists-collapse}). However, we note that in the subcritical state the onset of the exponential cutoff is much smaller than the system size. In contrast, in the supercritical state, the distributions show a sharp increase in the probability of large, system-spanning avalanches, as evidenced by the local maximum near the system size (SI, Fig.\ \ref{figS:avalanche-dists-collapse}).

\section{\label{sec:Ising} Maximum Entropy Modeling }   

The state of a neuron/electrode $i$ is represented by the binary variable $\sigma_i$. Therefore, the state of  our neural network or culture can be represented at each time step  by an $N-$dimensional variable $\bsigma = \{ \sigma_{1}, \sigma_{2}, ..., \sigma_{N} \}$. 
Let us denote by $\Pdata$ the probability of finding the system in one of the $2^N$ possible states.
The structure of $ \Pdata $ characterizes the properties of the system, but sampling all the $ 2^N $ possible states is infeasible even for moderately small networks. Alternatively, we can define $ \Pdata $ in a way that is consistent with a given set of $ \Nc \ll 2^{N} $  measurements, $f_m (\bsigma)$, whose expectation value $ \fcAve = \sum_{ \bsigma } f_m ( \bsigma ) \Pdata \approx \frac{1}{N_b} \sum_{k=1}^{N_b} f_m ( \bsigma(k) )$ for large $N_b$, where $ \sum_{ \bsigma } $ indicates a sum over all possible firing states $ \bsigma $. 
This amounts to finding a probability distribution $ \PMEM $ that maximizes the entropy  $\mathcal{ S } = - \sum_{ \bsigma } \PMEM \ln[ \PMEM ]$, subject to the constraints $ \fcAve = \sum_{ \bsigma } f_m ( \bsigma ) \PMEM $ \cite{nguyen_inverse_2017}.      
To solve this problem, we can use the method of Lagrangian multipliers \cite{jaynes_information_1957}. For each of the $ \Nc $ constraints $\langle f_m \rangle$, we have an associated Lagrangian multiplier $\lambda_m$, and an additional one, $\lambda_0$, is needed  to impose the normalization condition $\sum_{ \bsigma } \PMEM = 1$. The Lagrangian then reads, 
\begin{widetext}
\begin{equation} \label{eq:lagrangian}  
    \mathcal{ L } [ \PMEM ] = - \sum_{ \bsigma } \PMEM \ln[ \PMEM ] + \lambda_0 \cdot \left( \sum_{ \bsigma } \PMEM - 1 \right) + \sum_{m=1}^{ \Nc } \lambda_{m} \cdot \left( \sum_{ \bsigma } \PMEM f_m ( \bsigma ) - \fcAve \right) \textrm{ .}      
\end{equation}  
\end{widetext}     
Solving this problem involves finding the function $\PMEM$ for which the functional $\mathcal{ L } [ \PMEM ]$ attains an extremum. The solution is mathematically identical to a generalized Boltzmann distribution with temperature $T = 1$ (in units of the Boltzmann constant $k_B = 1$) \cite{tkacik_thermodynamics_2015,simoes_thermodynamic_2024} 
\begin{align}    
\label{eq:maxEntProbGeneral}    \PMEM &= \frac{ 1 }{ Z } e^{ -H ( \bsigma) } \textrm{ } \\       
\label{eq:maxEntProbZGeneral}    Z &= \sum_{ \bsigma } e^{ - H ( \bsigma ) } \textrm{ ,}  
\end{align}   
where one can recognize $ H ( \bsigma ) \equiv - \sum_{m=1}^{ \Nc } \lambda_{m} f_{m}(\bsigma) $ as a generalized Hamiltonian  and $Z$ as the corresponding generalized partition function. 
The next step consists in finding the ${ \lambda_m }$ that reproduce the measured expectation values from the data, which is known as the inverse Ising problem \cite{nguyen_inverse_2017}. In principle, each parameter $\lambda_m$ can be determined from the derivative of the logarithm of the partition function (\ref{eq:maxEntProbZGeneral}), $ \fcAve = \partial \ln[ Z ] / \partial \lambda_m $.  
However, solving this equation exactly becomes impractical when $N \gtrsim 20$,  as the number of terms in $Z$ grows exponentially as $2^{N}$. Alternatively, since we are trying to find the distribution $ \PMEM $ that best describes the empirical one, $ \Pdata $, we can impose that the set of parameters $\lambda_m$ minimize the so-called Kullback-Leibler divergence \cite{nguyen_inverse_2017} between these distributions,  
\begin{equation}                                             
    \Dkl \left( \Pdata || \PMEM \right) = \sum_{ \bsigma } \Pdata \ln \left( \frac{ \Pdata }{ \PMEM } \right) \textrm{ .}   
\end{equation}       
Performing the partial derivative of $\Dkl$ with respect to the parameters $\lambda_m$ gives 
\begin{equation} \label{eq:DklMinimization} 
    \frac{ \partial \Dkl }{ \partial \lambda_m } = \fcAveMEM - \fcAveData \textrm{ ,}  
\end{equation}
where $ \fcAveData \equiv \frac{1}{N_b} \sum_{k=1}^{N_b} f_m ( \bsigma(k) ) $ are the empirical averages measured from the data and $ \fcAveMEM \equiv \sum_{ \bsigma } f_m ( \bsigma ) \PMEM $ are the ones predicted by the ME distribution given by Eq.~(\ref{eq:maxEntProbGeneral}). 
Eq.~(\ref{eq:DklMinimization}) has two important implications.  
First, from the minimization condition $ \partial \Dkl / \partial \lambda_m = 0 $, the minimum of $\Dkl$ is reached when $\fcAveMEM = \fcAveData$, as intended. Second, it suggests a way to approach this minimum by updating each $\lambda_m$ proportionally to the corresponding difference $-(\fcAveMEM - \fcAveData)$.     
A possible method to reach this minimum is a gradient-descent method called Boltzmann Machine (BM) learning \cite{nguyen_inverse_2017, ackley_learning_1985}. This algorithm consists in sampling the Boltzmann distribution (\ref{eq:maxEntProbGeneral}) with a given  set of parameters $\lambda_m$, and estimating the averages $\fcAveMEM$ using a suitable method such as the Metropolis algorithm \cite{metropolis_equation_1953,bottcher_computational_2021}. Then,  the generated statistics, $\fcAveMEM$, are compared  with those from  data, $\fcAveData$, according to the following iterative scheme suggested by Eq. (\ref{eq:DklMinimization}) \cite{tkacik_spin_2009},  
\begin{equation} \label{eq:learningEq} 
    \lambda_m(n+1) = \lambda_m(n) - \BMLearningRate(n) \cdot \left( \fcAveMEM - \fcAveData \right) \textrm{ ,}   
\end{equation}      
where $n$ is the iteration index and $\BMLearningRate(n) \propto n^{-\alpha}$ is a decreasing learning rate, with the value of $\alpha > 0$ that can be adjusted depending on $N$ and the state of the dynamics of each dataset (SI, Table \ref{tableS:learning-rate-parameter}).               
   
\subsection{\label{subsec:k-pairwiseMEM} $K$-pairwise ME models } 
    
We are interested in finding the ME distribution $\PMEM$ that constrains the $N$ average local activities $\langle \sigma_i \rangle$, the $N \cdot (N - 1)/2$ pairwise correlations $\langle \sigma_i \sigma_j \rangle$, which in turn constrain  the correlation functions $C_{ij}$, and the  distribution of synchrony, $\PK$. 
This corresponds to $N$ Lagrange multipliers for $\langle \sigma_i\rangle$, denoted by $h_i$, $N \cdot (N - 1)/2$ for
$\langle \sigma_i\sigma_j\rangle$, denoted by $J_{ij}$, and $N+1$ for $\langle \delta_{K,K'(\bsigma)}\rangle$, denoted by $V_K$. Since only $N$  of the $N+1$ possible values of $P(K)$ are independent, we  only need to fit at most $N$ parameters $V_K$. Therefore, we set $V_{K=0}$ to zero. The Hamiltonian in (\ref{eq:maxEntProbGeneral}) then reads,   
\begin{equation} \label{eq:H}
    H ( \bsigma ) = - \sum_{i}^{N} h_i \sigma_i - \frac{1}{2} \sum_{i,j \neq i}^{N} J_{ij} \sigma_i \sigma_j - \sum_{K=0}^{N} V_K \delta_{K,K'(\bsigma)} \textrm{ ,}    
\end{equation}   
where $K'(\bsigma) = \sum_{i=1} ^ { N } \left( \sigma_i + 1 \right)/2$ counts the number of up-spins in configuration $\bsigma$, i.e. the number of firing neurons/electrodes in our case. Eq.~(\ref{eq:H}) is mathematically equivalent to the Hamiltonian of a generalized Ising model, known formally as a K-pairwise model \cite{tkacik_searching_2014,gardella_modeling_2019}, where $h_{i}$ is analogous to a local external field acting on spin $i$, $J_{ij}$ is an interaction constant between spins $i$ and $j$, and $V_{K}$ is a potential that depends only on the total magnetization $M(\bsigma) = 2K'(\bsigma) - N$. We note that  a higher value of $V_{K}$ indicates that the system favors states with $K$ up-spins. 

Each spin can model either the binary state of an IF neuron if the Hamiltonian parameters are inferred from IF model data, or the binary state of an electrode if the parameters are inferred from experimental data.      
The set of parameters $\{ h_{i} , J_{ij}, V_{K} \}$ is then learned using the BM algorithm, being updated at each iteration according to Eq.\ (\ref{eq:learningEq}), by applying simultaneously the following equations,   
\begin{align}
    \label{eq:hiUpdate}	 h_i(n+1) &= h_i(n) - \BMLearningRate(n) \cdot \left( \aveSigmaMEM - \aveSigmaData \right) \textrm{ ,} \\            
\label{eq:JijUpdate} J_{ij}(n+1) &= J_{ij}(n) - \BMLearningRate_{J}(n) \cdot \left( \aveTwoMEM - \aveTwoData \right) \textrm{ ,} \\  
\label{eq:VKUpdate} V_K(n+1) &= V_K(n) - \BMLearningRate(n) \cdot \left( \PKMEM - \PKData \right) \textrm{ ,}   
\end{align}  
where we set a slower learning rate $\BMLearningRate_J(n) = \BMLearningRate(n) / 2$ for the interaction constants $J_{ij}$ to avoid  impractical CPU times due to instabilities during the learning procedure,  since their number ($ \sim N^2 $) is much larger compared to the number of fields $h_i$ or potentials $V_K$ ($\sim N$).  
After $n = 5000$ iterations of the BM, the learning rate of the potentials $V_K$ is modified to $\BMLearningRate(n) \rightarrow \BMLearningRate(n) / \PKData$ which we heuristically found to improve the learning by enabling a more efficient convergence of the values for the smallest $P(K)$.  
 
We start with $h_{i} ( n = 1 ) = \aveSigmaData$, $J_{ij} ( n = 1 ) = 0$ and $V_{K} ( n = 1 ) = 0$ and then iterate equations (\ref{eq:hiUpdate})-(\ref{eq:VKUpdate}) typically until $n \sim 2 \cdot 10^5$. At each iteration, quantities are averaged over $M_c = 3 \cdot 10^5$ spin configurations using the Metropolis algorithm. We disregard the first $150N$ Monte Carlo iterations in order to reduce correlations with the initial state. To avoid divergence issues due to poor sampling at small $P(K)$, we only fit the $V_K$ associated with $P(K) > 10^{-4}$ for experimental data or $P(K) > 10^{-5} $ for the numerical data, and set all other $V_K = 0$. 
We use a smaller threshold for numerical data because $P(K)$ is better sampled  thanks to the larger number of time bins that can be considered---experimental data instead have a limited $\approx 1$h duration. 
A check for convergence is shown in the SI (Fig.\ \ref{figS:RMSD}).  The error functions associated with $\langle \sigma_i \rangle$, $\langle \sigma_i \sigma_j \rangle$, and $P(K)$,  defined as the root mean squared deviation between the empirical data and the Monte Carlo samples, eventually saturate close to zero, suggesting that our system is ergodic. 
At the end of the learning routine, we study the inferred Ising-like models with the set of fitted parameters $\{ h_{i}, J_{ij}, V_{K} \}$ by averaging over an increased amount of spin configurations $ M_c = 3 \cdot 10^6 $, averaged over $100$ random initial spin configurations, to reduce error bars.

\subsection{\label{subsec:parameters} Inferred parameters of the K-pairwise Ising models}

\begin{figure*}[!ht]       
	\includegraphics[width=0.9\linewidth]{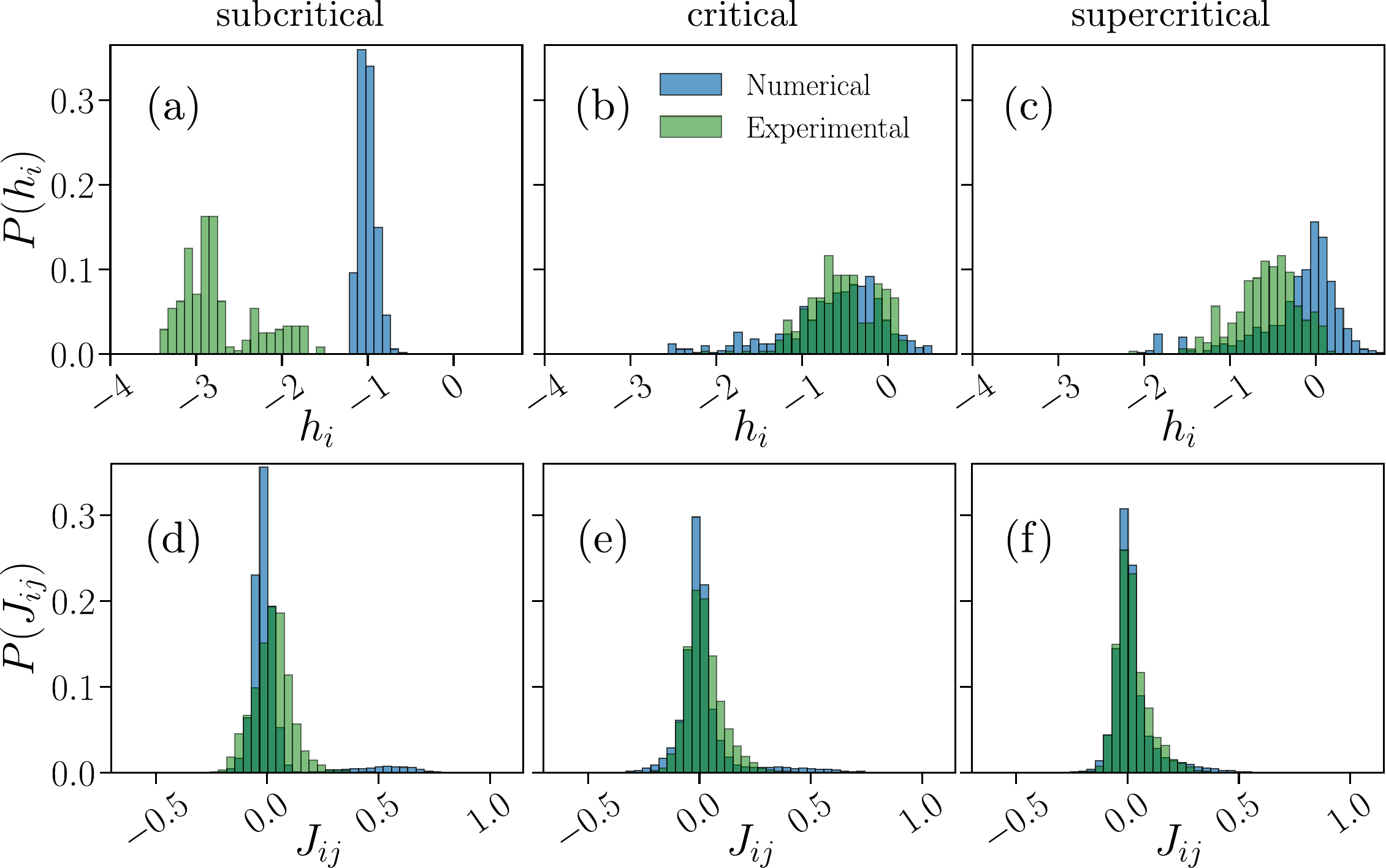}        
	\centering 
	\caption{\textbf{Fields and interaction constants of the K-pairwise Ising-like models inferred from experimental and numerical data in from subcritical to supercritical network states.} Distribution of the fields $h_i$ (a-c) and interaction constants $J_{ij}$ (d-f) obtained from the BM learning scheme (Eqs.\ (\ref{eq:hiUpdate})-(\ref{eq:VKUpdate})) of the generalized Hamiltonian (Eq.\ (\ref{eq:H})) for  which the probability distribution $\PMEM$ from Eq.~(\ref{eq:maxEntProbGeneral}) has expectation values consistent with the data of the average local activities $\aveSigmaData$, correlation functions $\CijData$ and synchronous probabilities $\PKData$, as measured in IF neural networks with $N=100$ neurons (blue histograms) and in neuronal cultures ($N = 60$) in the subcritical (left), critical (center) and supercritical (right) state, as well as experimental recordings of $N=60$ electrodes (green histograms). 
      }       
	\label{fig:parameters-hi-Jij}   
\end{figure*}

\begin{figure*}[!ht]         
	\includegraphics[width=0.9\linewidth]{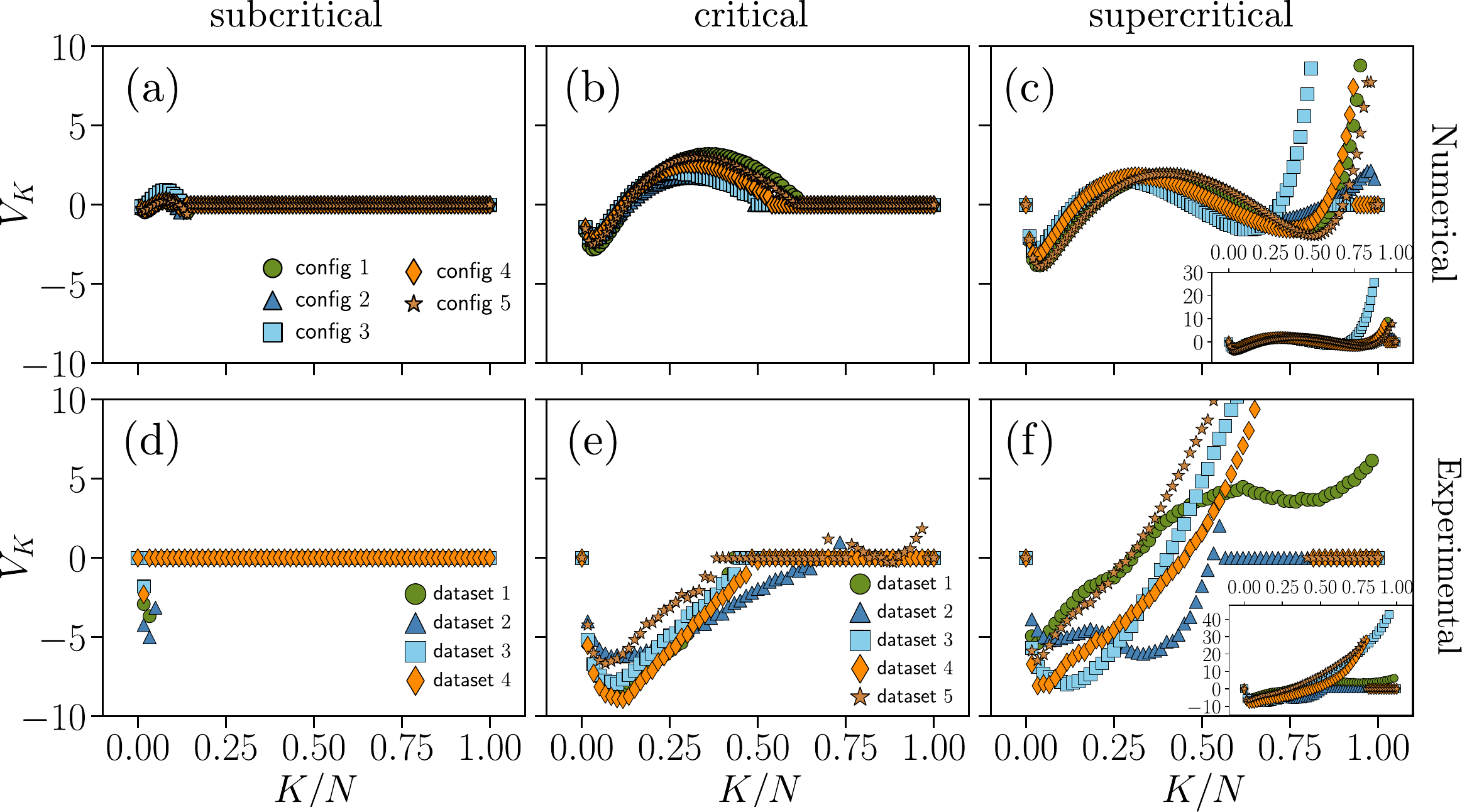}          
	\centering 
	\caption{\textbf{Potential functions of the K-pairwise Ising-like models inferred from the experimental and numerical data from subcritical to supercritical states.} Potential $V_K$ as a function of $K$ for numerical (a-c) and experimental (d-f) data in the subcritical (left), critical (center), and supercritical (right) state. Insets in (c, f): zoomed-out view of the main plot showing the full range of $V_K$.  
    }
	\label{fig:parameters-VK}   
\end{figure*}

In Figs.~\ref{fig:parameters-hi-Jij} and \ref{fig:parameters-VK}, we compare  the distributions of the fields $h_i$, interaction constants $J_{ij}$, and the potential $V_K$ as a function of $K$, obtained from the BM learning process on the IF model and experimental data. 
Each set of parameters reproduces the original input data for the corresponding IF model network ($N=100$ neurons) or neuronal culture ($N=60$ electrodes) and the corresponding dynamical regime (subcritical, critical, or supercritical). 
For the quality of the fit, we refer the reader to the SI, Figs.\ \ref{figS:quality-of-fit-critical}-\ref{figS:quality-of-fit-experimental-supercritical}, which provide an assessment of convergence for $h_i$, $J_{ij}$, and $V_K$ based on the similarity between the respective empirical averages $\langle \ldots \rangle _{\mathrm{data}}$ and the Monte Carlo averages $\langle \ldots \rangle _{\mathrm{MEM}}$.  

For the IF model, the fields $h_i$ (Fig.~\ref{fig:parameters-hi-Jij}a-c, blue histograms) are overall negative in the subcritical and  critical states, with an increasing probability of positive values when moving towards the supercritical state. In the subcritical state, the $h_i$ are in a very narrow range around -1,  which implies that  neurons fire very sparsely, consistently with the distribution of $\langle \sigma_i \rangle$ (SI, Fig.\ \ref{figS:aveSigma-dists}a). In the critical state instead, the distribution of $h_i$ broadens and shifts towards zero, with non-zero probabilities for small positive $h_i$ (Fig.\ \ref{fig:parameters-hi-Jij}b, blue histograms), showing an increasing heterogeneity in the local fields. This indicates that, at criticality, neuronal firing is still sparse, but the network tends to be more active (see SI, Fig.\ \ref{figS:aveSigma-dists}). Indeed, the average firing rate increases by about ten times  compared to the subcritical state (Fig.\ \ref{fig:allometry}). When moving towards the supercritical state, the probability of positive $h_i$ increases significantly, signaling a consistent increase in firing activity with respect to the baseline critical state (see also SI, Fig.\ \ref{figS:aveSigma-dists}a), the average firing rate being about ten times larger than in the critical state (Fig.\ \ref{fig:allometry}). 

The results for our IF model closely recapitulate the distributions $P(h_i)$ learned from the experimental data (Fig.~\ref{fig:parameters-hi-Jij}a-c, green histograms). The hypoexcitable cultures (AP5/DNQX), which were classified as subcritical  on the basis of avalanche metrics \cite{shew_neuronal_2009}, consistently show negative $h_i$ only, distributed between -2 and -3 (Fig.~\ref{fig:parameters-hi-Jij}a, green histograms). This scenario is consistent with the strongly reduced firing rate in these cultures \cite{shew_neuronal_2009} (SI, Fig.\ \ref{figS:aveSigma-dists}d). In cultures treated with AP5 only (SI, Fig.\ \ref{figS:AP5-results}) the $h_i$ tend to be less negative and mostly close to -1. We note that AP5 acts as an NMDA receptor antagonist, suppressing the slower, more sustained form of excitatory transmission between neurons. DNQX, on the other hand, blocks AMPA receptors, which mediate fast excitatory signaling. Although AP5-treated cultures retain many characteristics of critical dynamics, only the combined inhibition of NMDA and AMPA receptors using AP5 and DNQX effectively shifts the cultures into a subcritical state by reducing their overall excitability. 
It is important to notice that, unlike model simulations, in this case neural dynamics tend to show more variability across samples (i.e.\ cultures), also in terms of distance from criticality \cite{shew_neuronal_2009}.  In the baseline critical cultures, we find that $P(h_i)$ is remarkably close to the one inferred from numerical data, covering a range of values between -1 and 0 that is consistent with the observed increase in firing activity \cite{shew_neuronal_2009} (SI, Fig.\ \ref{figS:aveSigma-dists}e). We observe a similar distribution of $h_i$ in the disinhibited cultures treated with PTX and classified as supercritical \cite{shew_neuronal_2009}, which however present an increased firing rate as compared to the critical case (SI, Fig.\ \ref{figS:aveSigma-dists}f).

The second set of parameters inferred from numerical and experimental data provides the interaction constants $J_{ij}$ of the Hamiltonian (Eq.\ \ref{eq:H}).
In both numerical (Figs.~\ref{fig:parameters-hi-Jij}d-f, blue histograms) and experimental data (Figs.~\ref{fig:parameters-hi-Jij}d-f, green histograms), the distribution of $J_{ij}$ is centered around $J_{ij} \approx 0$ for all network states. We observe little or no  modulations in $P(J_{ij})$ across network states, in particular between the critical and supercritical  state,   both in experimental and numerical data. Notably, subcritical IF networks display a small subset of positive interaction constants (Fig.~\ref{fig:parameters-hi-Jij}d, blue histograms) around $J_{ij} \approx 0.5$. This mild bimodal behavior is further attenuated in the critical state, and is not present in the supercritical case, where  the distribution becomes  unimodal  with a  heavy positive tail as for experimental data.   

Finally, we examine the inferred potentials $V_K$ (Fig.\ \ref{fig:parameters-VK}), which are related to the distribution of synchrony, $P(K)$ (Fig.\ \ref{fig:PK}). Here, we first discuss the results for our IF model, and then compare them with the experimental data. In the  subcritical state (Fig.~\ref{fig:parameters-VK}a), the $V_K$ inferred from the model are approximately zero for all $K$, indicating  that  synchronous firing of even  a small fraction of neurons is not likely in this state---as demonstrated by the distribution $P(K)$ (Fig.~\ref{fig:PK}a) and in line with evidence that collective bursts are rare in weakly excitable networks \cite{shew_neuronal_2009}. This may make global observables such as the synchronous probability $P(K)$ less relevant in this regime.    

Both the critical and supercritical states present a markedly different scenario.  At criticality (Fig.~\ref{fig:parameters-VK}b), the $V_K$ have a negative minimum at low values of $K/N = 0.03\pm0.01$, where the uncertainty is estimated as the range between the minimum and maximum values obtained from the different datasets, and are positive over a range of  intermediate $K$---with a maximum at $K/N = 0.35 \pm 0.10$---,  which would favor synchronous co-activation of neurons consistent with transient collective bursts of activity. Indeed, in the critical state, the $P(K)$ is non-zero over a broad range of $K$ (Fig.~\ref{fig:PK}b). Such features of the potentials $V_K$ persist in the supercritical states (Fig.~\ref{fig:parameters-VK}c), and become slightly more pronounced, in particular the  minimum at $K/N = 0.03\pm0.01$. At the same time, the supercritical state shows a key distinctive feature in most numerical samples, namely very high values of $V_K$ for $K/N > 0.75$. Consistently, the $P(K)$ has  a  much longer tail (Fig.~\ref{fig:PK}c). This  is closely  related to the sharp increase in the probability of very large avalanches in the supercritical regime (SI, Fig.\ \ref{figS:aval-dists}), and is consistent with strong co-activation of neuronal populations in disinhibited networks \cite{shew_neuronal_2009,lombardi_temporal_2016}.

Comparison with $V_K$ inferred from cultures (Fig.~\ref{fig:parameters-VK}d-f) shows some common features and some important differences, particularly in the critical state. In cultures treated with AP5 and DNQX, $V_K$ is always zero, except for a localized negative value at $K/N = 0.02 \pm 0.01$. This is in line with inference from our network model and reflects the absence of collective firing  in hypoexcitable cultures \cite{shew_neuronal_2009}. Accordingly, the distributions $P(K)$ show a sharp exponential decay (Fig.~\ref{fig:PK}d). All these cultures were originally classified as subcritical and showed no power-law behavior in avalanche size and duration distributions \cite{shew_neuronal_2009}.    

An important exception is represented by the one culture that was treated only with AP5 (SI, Fig.\ \ref{figS:AP5-results}). Although classified as subcritical, the $V_K$ are strongly negative over a wider range of $K$ (SI, Fig.~\ref{figS:AP5-results}f), and $P(K)$ exhibits a broad tail (SI, Fig.~\ref{figS:AP5-results}c). 
This scenario closely resembles the behavior of $V_K$ at criticality, where negative non-zero values characterize $V_K$ for $K/N < 0.5$, with a minimum $K/N = 0.10 \pm 0.03$, and $V_K$ is always zero for $K/N > 0.5$ (Fig.~\ref{fig:parameters-VK}e). This confirms that only the combined inhibition of NMDA and AMPA receptors using AP5 and DNQX effectively produces a subcritical network state.
We notice that,  unlike in  our model, $V_K$ is almost never positive in cultures  at criticality.  
Yet, the distribution of synchrony, $P(K)$, for critical cultures qualitatively matches our model at criticality (Figs.~\ref{fig:PK}b, e).  

Moving to the supercritical state (PTX),  the $V_K$ remain mostly negative for $K/N < 0.5$ in all cultures (local minimum $K = 0.15 \pm 0.15$). On the contrary, most cultures show increasing positive $V_K$ for $K/N > 0.5$ (Fig.~\ref{fig:parameters-VK}f). Importantly, we observed a similar behavior in our network model (Fig.~\ref{fig:parameters-VK}c). As in the model, supercritical cultures are characterized by higher firing rates, a very broad  distribution of neural synchrony (Fig.~\ref{fig:PK}f), and an excess of large avalanches \cite{shew_neuronal_2009}.

Overall, these results show that our neural network model is well described by a generalized Ising-like model that closely resembles the Ising-like model inferred  from neuronal data classified as either subcritical, critical or supercritical by means of avalanche-based metrics. Importantly, our model  also captures the main features of avalanche size and duration distributions at and away from criticality (SI, Fig.~\ref{figS:aval-dists}).

\subsection{\label{subsec:prediction} Predictive capability of the K-pairwise Ising models}

\begin{figure*}[ht]   
	\includegraphics[width=0.9\linewidth]{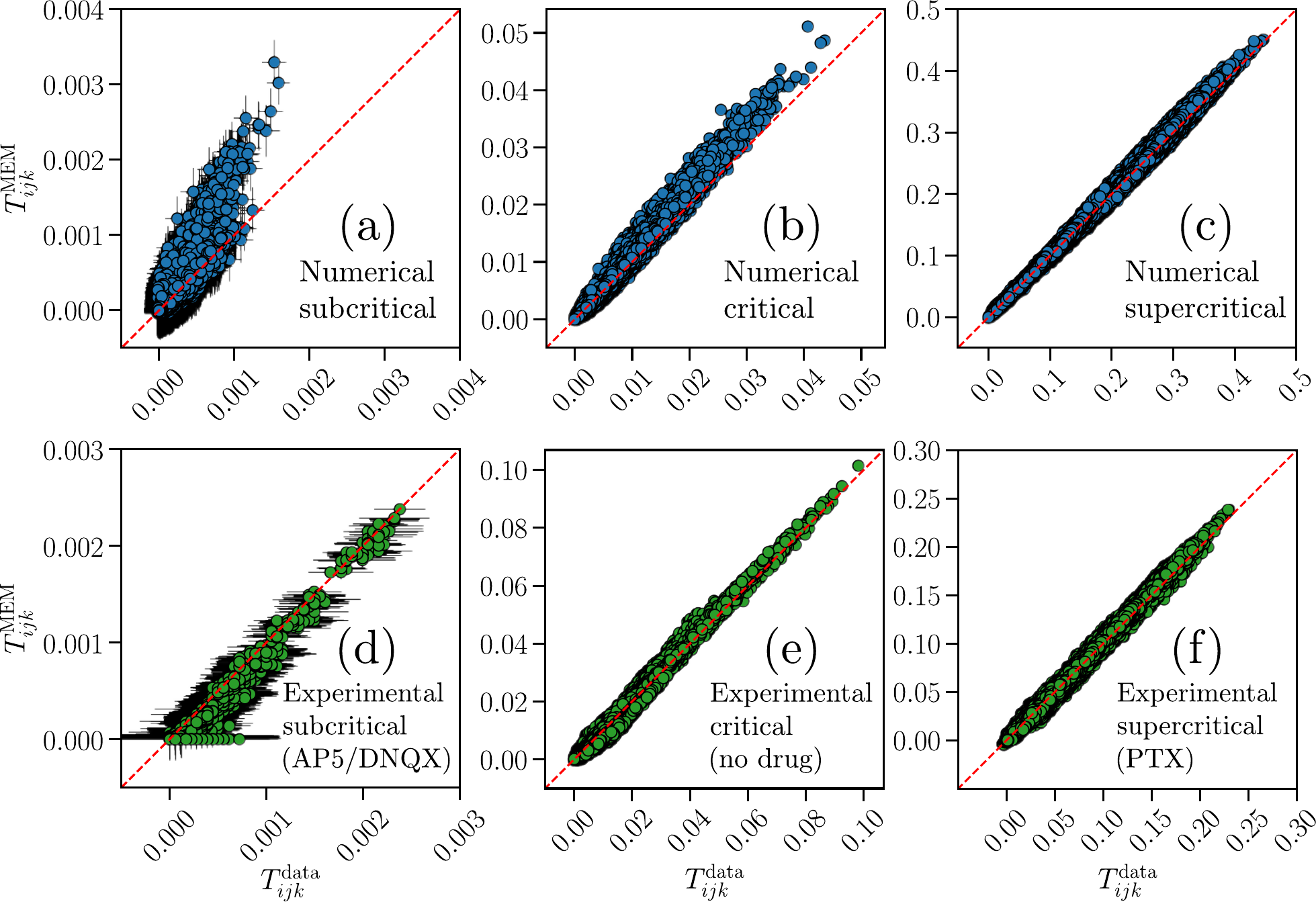}       
	\centering 
	\caption{\textbf{Predictive capability of the K-pairwise Ising models inferred from numerical and experimental data.} Comparison of the three-point correlation functions $T_{ijk}$ between the data (x-axes) and the prediction of the maximum entropy distribution (Eq.\ (\ref{eq:maxEntProbGeneral})) (y-axes), for IF neural networks with $N=100$ neurons and \textit{in vitro} cultures ($N = 60$)  in the subcritical (a, d), critical (b, e), and supercritical (c, f) state. 
    Red dashed lines indicate the bisector $y = x$. For numerical data (x-axes), results are averages over $N_b = 10^{7}$ time bins; $N_b \approx 1.4 \cdot 10^{5}$ for the experimental data. For the sampling of the distribution $\PMEM$ from Eq.~(\ref{eq:maxEntProbGeneral}) (y-axes), using the respective learned parameters shown in Figs.\ \ref{fig:parameters-hi-Jij} and \ref{fig:parameters-VK}, results are averages over $M_c = 3 \cdot 10^{6}$ spin configurations, and over $100$ random initial spin configurations. Error bars represent the standard error of the mean and  are smaller than or equal to the symbol size, with the exception of  (a) and (d).
    }
	\label{fig:predictive-capability-triplets}    
\end{figure*}

The predictive capability of the generalized Ising models can be tested by comparing quantities not constrained by the ME modeling scheme, such as the three-point correlation functions $T_{ijk}$, 
 
\begin{widetext}  
\begin{align} \label{eq:triplets} 
   	T_{ijk} = \lchev \left( \sigma_i - \lchev \sigma_i \rchev \right) \cdot \left( \sigma_j - \lchev \sigma_j \rchev \right) \cdot \left( \sigma_k - \lchev \sigma_k \rchev \right) \rchev = \lchev \sigma_i \sigma_j \sigma_k \rchev   
						- \lchev \sigma_i \rchev \lchev \sigma_j \sigma_k \rchev 
						- \lchev \sigma_j \rchev \lchev \sigma_i \sigma_k \rchev  
						- \lchev \sigma_k \rchev \lchev \sigma_i \sigma_j \rchev + 2 \lchev \sigma_i \rchev \lchev \sigma_j \rchev \lchev \sigma_k \rchev \textrm{ .}  
\end{align}   
\end{widetext}   

In Fig.\ \ref{fig:predictive-capability-triplets} we compare the three-point correlation functions predicted by the ME distribution, $\TijkMEM$ (y-axes),  with those from the original data, $\TijkData$ (x-axes), for both IF model networks and neuronal cultures, in the subcritical, critical, and supercritical states. We find that the generalized Ising models reproduce the three-point correlations, $T_{ijk}$, rather accurately in the critical and supercritical states (Fig.~\ref{fig:predictive-capability-triplets}b-c)  where $T_{ijk}$ are stronger (Fig.~\ref{fig:predictive-capability-triplets}e-f). Predictions are less accurate in the subcritical state, particularly for the numerical model (Fig.~\ref{fig:predictive-capability-triplets}a). We note that three-point correlations are extremely weak in the subcritical state.  In both the network model and  cultures treated with AP5/DNQX, we find $T_{ijk} < 10^{-3}$---more than one order of magnitude smaller than in critical and supercritical states (Fig.~\ref{fig:predictive-capability-triplets}a and Fig.~\ref{fig:predictive-capability-triplets}d). In contrast,  for the reasons discussed above, the culture treated with AP5 shows three-point correlations similar to cultures at criticality (SI, Fig.\ \ref{figS:AP5-results}g), despite the fact that avalanche analysis suggests that it is in a subcritical state \cite{shew_neuronal_2009}.

Overall,  we find that the Ising-like models predict $T_{ijk}$ more accurately for the experimental data than for the neural network model. In particular,  the inferred Ising-like models tend to substantially overestimate $T_{ijk}$ for the network model in the subcritical state (Fig.~\ref{fig:predictive-capability-triplets}a).  
This may be due to the extremely small values of the $T_{ijk}$ combined with larger errors in the fitted features that  are used to obtain the estimates of $T_{ijk}$. Indeed, propagation of small errors on the fitted features may lead to the observed mismatches, which are $< 10^{-3}$.

\subsection{\label{subsec:thermodynamics} Thermodynamics of the K-pairwise Ising models} 
  
\begin{figure*}[ht]   
	\includegraphics[width=\linewidth]{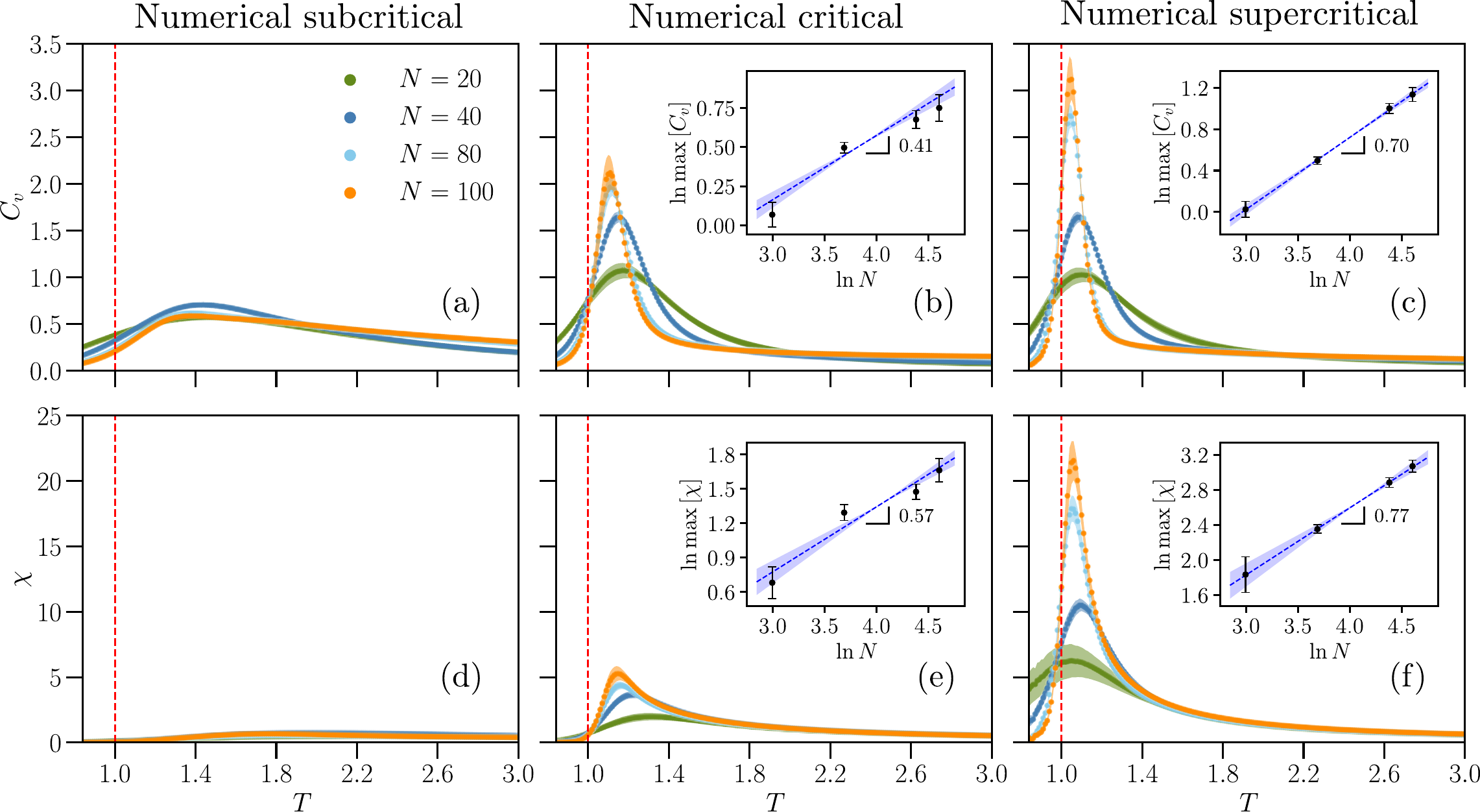}       
	\centering 
	\caption{\textbf{Thermodynamic response functions of the K-pairwise Ising-like models inferred from numerical data.} Specific heat $C_v$ (a–c) and susceptibility $\chi$ (d–f) as functions of temperature $T$ for K-pairwise Ising models fitted to IF networks of size $N \in \{20, 40, 80, 100\}$ in subcritical (left column), critical (center column), and supercritical states (right column). Vertical dashed lines indicate $T = 1$, where the $T$-parameterized probability (Eq.\ (\ref{eq:PTemp})) matches the maximum entropy distribution (Eq.\ (\ref{eq:maxEntProbGeneral})) that fits the data. For each $T$, results are averaged over $M_c = 3 \cdot 10^6$ spin configurations, and $100$ random initial spin configurations. 
    The colored shaded areas around the curves of $C_v$ and $\chi$ represent the standard error obtained from $5$ different IF network configurations for each $N$. The insets show the maxima of $C_v$ (b,c) and $\chi$ (e,f) as a function of $N$ (log–log scale), with error bars representing the standard error across 5 IF network configurations. Dashed lines are linear least-squares fits of the form  $\log(\mathrm{max}\left[ C_v \right]) = A\cdot \log N + B$ and $\log(\mathrm{max}\left[ \chi \right]) = A\cdot \log N + B$. Blue shade indicates  the range spanned by individual fits and corresponds to slope $\pm$ standard error,  estimated by bootstrapping.
    }
	\label{fig:thermodynamics}   
\end{figure*}           

\begin{figure*}[ht]    
	\includegraphics[width=\linewidth]{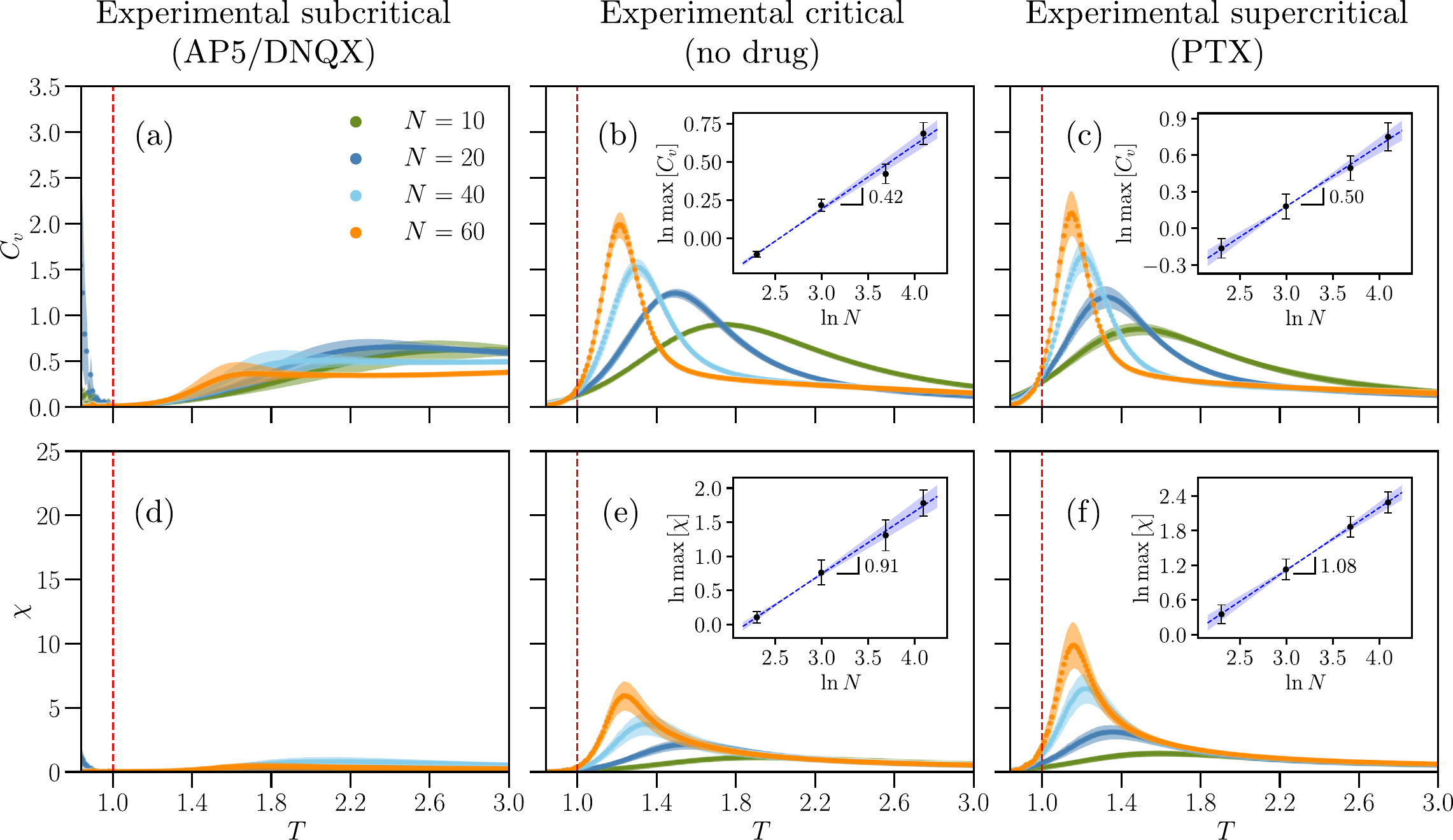}        
	\centering 
	\caption{\textbf{Thermodynamic response functions of the K-pairwise Ising-like models inferred from experimental data.} Same as in Fig.\ \ref{fig:thermodynamics} but for  data recorded in cortex slice cultures ($N = 60$ total electrodes; random subsampling at sizes $N \in \{10,20,40\}$) treated with a combination of AP5/DNQX (left column), in baseline, no-drug cultures (center column) and in cultures treated with PTX (right column). Shaded areas around the curves of $C_{v}$ and $\chi$ correspond to the standard error obtained from $5$ (b, c, e, f) or $4$ (a, d) experimental samples.
    }        
	\label{fig:thermodynamics-experimental}    
\end{figure*}           
 
The ME distribution (\ref{eq:maxEntProbGeneral}) can be generalized by introducing an additional parameter $T$ \cite{tkacik_thermodynamics_2015}, 
\begin{equation} \label{eq:PTemp}  
    P(\bsigma,T) \equiv \frac{ 1 } { Z(T) } e ^ { - H(\bsigma) / T } \textrm{ ,}   
\end{equation} 
with $Z(T) = \sum_{\bsigma} e^{-H(\bsigma)/T}$. Here, $T$ is a temperature-like parameter in units of the Boltzmann constant ($k_{B} = 1$), which sets the strength of the Ising parameters $\{ h_i, J_{ij}, V_{K} \}$ by uniformly rescaling them by the same factor $1/T$. This parameter is useful for probing thermodynamic properties of the models, and for investigating whether they exhibit notable behaviors near $T=1$, where $P(\bsigma,1) = \PMEM$.  

Given a set of $N$ spins, the total magnetization $M(\bsigma) = \sum_{i=1}^{N} \sigma_i$ and the energy $H(\bsigma)$ defined in Eq.~(\ref{eq:H}) can be obtained for different temperatures. From the fluctuations of these quantities, according to the fluctuation-dissipation theorem, the isothermal magnetic susceptibility $\chi$ and specific heat $C_v$ can be calculated as functions of $T$, 
\begin{align}      
\label{eq:sus}	\chi &= \frac{1}{T N} \cdot \left( \left\langle M^{2} \right\rangle - \left\langle M \right\rangle ^{2} \right)  \textrm{ ,} \\ 
\label{eq:cv}	C_v &= \frac{1}{T^2 N} \cdot \left( \left\langle H^2 \right\rangle - \left\langle H \right\rangle ^2 \right) \textrm{ ,}     
\end{align}    
where $\langle \cdots \rangle$ indicates an average over spin configurations generated by sampling Eq.~(\ref{eq:PTemp}) using Monte Carlo simulations.    
      
In Figs.~\ref{fig:thermodynamics} and \ref{fig:thermodynamics-experimental}, we show the Monte Carlo results for $\chi$ and $C_{v}$ as functions of the  temperature $T$ in the K-pairwise Ising models inferred from the IF neural network and from cultures at and off criticality. For the Ising-like model inferred from the IF networks, we consider several system sizes, $N \in \{20, 40, 60, 100\}$. For the cultures, we considered $N \in \{10, 20, 40, 60\}$, with $N = 60$ being the full set of recording electrodes. Smaller $N$ were obtained by random subsampling from the full set of  electrodes. We verified that results are independent of the initial spin configurations of the Monte Carlo simulations (SI, Figs. \ref{figS:thermodynamics-IC-comparison} and \ref{figS:thermodynamics-IC-comparison-experimental}).   

The K-pairwise Ising models inferred from  real and IF networks in the critical state exhibit maxima in the specific heat and susceptibility at $T = \Tmax$ slightly larger than one (see SI, Table~\ref{tableS:Tmax} for the values of $\Tmax$). As the system size increases, $\Tmax$ consistently approaches the unit temperature,  while the maxima in $C_v$ and $\chi$ increase  (Figs.~\ref{fig:thermodynamics}b, e and \ref{fig:thermodynamics-experimental}b, e), growing approximately as  $\maxCv \propto N^a$ with $a=0.41\pm0.06$ (IF networks) and $a=0.42\pm0.04$ (random subsampling of real networks), and $\maxChi \propto N^b$ with $b=0.57\pm0.09$ (IF networks) and $b=0.91\pm0.11$ (random subsampling of real networks). 
This suggests  that in the thermodynamic limit $N \to \infty$ the inferred Ising models operate at a critical point, which is consistent with the classification of the network state based on avalanche metrics.
 
Importantly, the models inferred from supercritical IF (Fig.~\ref{fig:thermodynamics}c, f) and real neuronal  networks (Fig.~\ref{fig:thermodynamics-experimental}c, f) also display maxima in $C_v$ and $\chi$ close to $T=1$, with $\Tmax$ approaching one for increasing system sizes. In the IF networks (Fig.~\ref{fig:thermodynamics}c, f), the maximum of $C_v$ increases with the number of  neurons, $N$, faster than in the critical case, i.e.\ $\maxCv \propto N^a$ with $a=0.70\pm0.06$. Unlike $C_v$, the maximum of $\chi$ follows a scaling consistent with the critical case, i.e.\ $\maxChi \propto N^b$ with $b=0.77\pm0.12$. In real supercritical networks instead (Fig.~\ref{fig:thermodynamics-experimental}c, f), maxima in $C_v$ and $\chi$ scale with $N$ with exponents consistent with those observed at criticality ($a=0.50\pm0.07$ vs $a=0.42\pm0.04$;  $b=1.08\pm0.13$ vs $b=0.91\pm0.11$). As a word of caution, we note that the reported power-law fits for the specific heat and susceptibility as a function of $N$ should not be understood as a finite size scaling (FSS) analysis, but merely as a way to put on a quantitative ground the comparison between the critical and supercritical neural networks. A proper and reliable FSS analysis of these models would indeed require scaling up (or down) the system size—not equivalent to subsampling from the population of neurons—, with the possibility of accessing larger system sizes. Moreover, learning ME models with an adequately large number of units is still prohibitive, due to its computational costs. 

In contrast to models inferred from critical and supercritical networks, models inferred from subcritical networks do not show clear maxima in $C_v$ and $\chi$ (Figs.~\ref{fig:thermodynamics}a, d and Figs.~\ref{fig:thermodynamics-experimental}a, d), in particular those inferred from subcritical cultures. Furthermore, $C_v$ and $\chi$ do not increase with the system size  (Figs.~\ref{fig:thermodynamics}a, d and \ref{fig:thermodynamics-experimental}a, d).

Recent analyses of neural network models have found that when ME  models are fit to highly correlated subsets of neurons, the resulting model appears more critical than it actually is \cite{nonnenmacher_signatures_2017}. In \cite{nonnenmacher_signatures_2017}, neurons were correlated only through common stimuli, so it is unclear whether this effect persists in networks of recurrently connected neurons. We tested this on our model fits of the subcritical networks and found that, if we bias subsampling towards the most active and  correlated neurons in the IF network, maxima in both $C_v$ and $\chi$ are, to some extent, enhanced compared to the random subsampling case (SI, Fig.~\ref{figS:thermo-subsampling}). This indicates that subsampling of neural systems, which is common, may bias inference of the network state.  

The combination of AP5 and DNQX inhibits  NMDA and AMPA receptors, driving cultures towards the above-described subcritical scenario. In contrast, cultures treated only with AP5 exhibit maxima in $C_v$ and $\chi$ that are consistent with those observed in the critical state (compare Fig.\ \ref{fig:thermodynamics-experimental}a,d to SI Fig.\ \ref{figS:AP5-results}h,i). This indicates that  simultaneous inhibition of NMDA and AMPA receptors is necessary to drive the cultures into a subcritical state.

\section{\label{sec:Discussion} Discussion}       
 
In this study we presented a thorough analysis of static ME models inferred from cultures of neurons at and away from criticality, and compared them with equivalent models inferred from  integrate-and-fire neural networks  that can be tuned to operate at and away from criticality. 
We assumed an operational definition of criticality based on neuronal avalanche metrics, as originally defined for cultures of neurons \cite{shew_neuronal_2009}. The IF model correctly reproduces these metrics at criticality, i.e.\ exponents of power-law size and duration distributions, shows absence of scaling in the subcritical state (exponential-like distributions), and shows a sharp increase in large avalanches (order of the system size) in the supercritical state. Here, we showed that, despite the intrinsic difference in the underlying dynamics, ME models inferred from this IF neural network closely match those that are inferred from neuronal  data, in particular the neuronal coupling structure and the local fields $h_i$ (see distributions of $J_{ij}$ and $h_i$ in Fig.\ \ref{fig:parameters-hi-Jij}). 

Importantly, this close equivalence extends to thermodynamic quantities such as the specific heat, $C_v$, and the susceptibility  $\chi$ (Figs.\ \ref{fig:thermodynamics} and \ref{fig:thermodynamics-experimental}). In Ising models inferred both from simulation and experimental data at criticality, $C_v$ and $\chi$ show pronounced maxima near the effective temperature $T = 1$---the temperature at which ME models were inferred. As expected at criticality, these maxima increased with the system size in IF networks and with the subsampling size in real cultures. However, our results showed that such maxima  persist in systems (numerical and experimental) that are classified as supercritical on the basis of avalanche metrics. In contrast, ME models inferred from subcritical systems, both numerical and experimental, do not show such evidence of criticality. This indicates that static ME models, which do not take into account dynamical properties, correctly distinguish between systems classified as subcritical and critical/supercritical according to neuronal avalanche metrics. However, they may not be able to discriminate between avalanche criticality and supercriticality, although they may still capture a number of important features, as we shall discuss in turn.

\subsection{Modulation of ME modeling parameters across network states} 

The local fields $h_i$ are key parameters in the inferred Ising models that control the excitability of neurons. We observed that the distribution of $h_i$ shifts from strongly negative values toward less negative and positive values as systems  are driven from the subcritical  to the critical and  supercritical state. This trend  reflects, to some extent, the increase in   average firing rate observed when moving from subcritical to supercritical states, both in our network model and in neural data \cite{shew_neuronal_2009}. In neural data, a subcritical state is induced by reducing  network excitability, whereas supercritical states are obtained by reducing inhibition. Both interventions alter the excitation/inhibition balance of the network, which not only affects neural firing rates, but also has a strong impact on collective, synchronous firing. This is demonstrated  by the broadening of  the distribution of synchrony $P(K)$ when moving towards the critical and supercritical states (Fig.\ \ref{fig:PK}), and  also reflected in avalanche size and duration distributions \cite{shew_neuronal_2009,lombardi_avalanche_2019}.
 
A similar behavior is found in our neural network model. In the K-pairwise ME modeling, one constrains the distribution $P(K)$, which provides at most $N$ additional potentials $V_K$. When  there is little  or no synchronous firing across the network, as in the subcritical state, $V_K$ are mostly zero. On the contrary, $V_K$ are not negligible in critical and supercritical states. Notably, we found that $V_K$ are strongly altered in the transition from criticality to supercriticality, in particular at large $K$, where they become strongly positive---in the same range where $V_K$ are zero at criticality. Among all the inferred parameters, the fields $V_K$ show the most striking  difference when comparing critical and supercritical states. Because they are related to a collective variable, the modulations  in $V_K$  can be easily translated into changes in avalanche dynamics, as outlined above.

We observed similar  modulations of $V_K$ across network states for the neural network model and the experimental data. However, we note an important difference between the two. While $V_K$ are always negative or zero in data-inferred Hamiltonians and only become positive at large $K$ in the supercritical state, in model-inferred Hamiltonians they are mostly positive or zero. This difference may arise from  the  model   dynamics  controlling the emergence of synchronous firing, which are likely to differ from those  that underlie neural activity in cultures. 

Comparing our results with \cite{simoes_thermodynamic_2024}, 
we note that including the constraint on the probability of synchronous firing  improves the predictive capabilities of the Ising-like model for higher-order correlations, as also reported in \cite{tkacik_searching_2014}.  Indeed, in \cite{simoes_thermodynamic_2024}, the $P(K)$ was not constrained and the inferred model systematically overestimated higher-order correlations at criticality (compare Fig.~\ref{fig:predictive-capability-triplets}b with Fig.\ 6 in \cite{simoes_thermodynamic_2024}). 
 
\subsection{Thermodynamic quantities across network states} 

We have shown that  K-pairwise Ising models inferred from critical and supercritical systems  show pronounced maxima in specific heat $C_v$ and susceptibility $\chi$  near the effective temperature $T=1$. This feature is common to both the network model and the neural cultures  (Figs.~\ref{fig:thermodynamics}b-c, f-g and \ref{fig:thermodynamics-experimental}b-c, f-g).   
Moreover, we also demonstrated that these maxima grow superlinearly with the number of neurons in our network model and with the number of sampled electrodes in cortex slice cultures, both in the  critical and supercritical states. Overall, this evidence would  suggest that, in both cases, the system is at or close to criticality. However, this would contradict their classification based on avalanche dynamics.  A similar contradiction emerges  when we consider an individual culture treated with AP5 only,  which acts as an NMDA receptor antagonist and  suppresses  slower and more sustained form of neuronal excitation.  For this culture, both $C_v$ and  $\chi$ showed pronounced maxima near $T = 1$, suggesting instead a critical state (SI, Fig.~\ref{figS:AP5-results}h, i), and indicating that only combined inhibition of NMDA and AMPA receptors drives the cultures into a subcritical state by reducing their overall excitability. 

Criticality in neuronal systems  has been hypothesized to optimize stimulus response and  maximize function such as dynamic range, the range of stimuli that can be processed by the network to produce a functional response, information storage and capacity \cite{shew_information_2011}. In \cite{shew_neuronal_2009}, it was shown that  cultures at criticality have a much higher dynamic range compared to  sub- and supercritical cultures. The dynamic range is related to the responsiveness of the network  to external stimuli, i.e.\ to its susceptibility. Thus, we would expect the susceptibility to be high  for Ising models inferred from critical cultures and low for  models inferred from sub- and supercritical cultures. Although this prediction is met for cultures treated with AP5/DNQX, classified as subcritical (except for the one treated with AP5 only), we observe a pronounced maximum in the susceptibility  for supercritical cultures.

\subsection{Identifying criticality in neural data} 

While showing a certain correspondence between dynamical  criticality (presence of power-law in neuronal avalanche statistics) and static criticality (maxima in $C_v$ and $\chi$ near the unit temperature in the inferred Ising-like models), our analysis  points to important differences, particularly when trying to assess deviations from the critical state in neuronal populations.
Therefore, caution must be taken when drawing  conclusions about criticality or deviations from criticality based on static ME modeling approaches, which do not constrain dynamical properties. In particular, we note that temporal dynamics are strongly altered in PTX-induced supercritical states \cite{lombardi_temporal_2016,lombardi_avalanche_2019}.  
Thus, ME approaches that incorporate dynamic information, e.g.\ the joint distribution of the number of spiking neurons in different time windows \cite{mora_dynamical_2015} or the temporal correlations \cite{cavagna_dyn_ME_2014}, may be more suitable to distinguish between critical and supercritical states, as also  suggested by recent analysis  of neuronal models \cite{serafim_maximum-entropy-based_2024}. 

Alternative approaches, e.g.\ based on renormalization group (RG) ideas \cite{meshulam_prl_2019},  or definitions of criticality can be used,  which do not necessarily correspond to simultaneous emergence of neuronal avalanche scaling. An example of this kind is provided by edge-of-chaos criticality \cite{kanders_chaos_2017,dhamen_pnas_2019}. On the other hand, recent work indicates that neuronal avalanches and non-trivial RG scaling both emerge in the resting-state of the human brain in a narrow, slightly subcritical dynamical region \cite{topal_RG_2026}.  However, how RG-based measures, e.g.\ scaling exponents, are related to the avalanche-based  classification of network states remains to be clarified, in particular for the supercritical regime.

\begin{acknowledgments}   

L.d.A. acknowledges support from the Italian MUR project PRIN2017WZFTZP and from NEXTGENERATIONEU (NGEU) funded by the Ministry of University and Research (MUR), National Recovery and Resilience Plan (NRRP), and project MNESYS (PE0000006)-A multiscale integrated approach to the study of the nervous system in health and disease (DN. 1553 11.10.2022). H.J.H. thanks the INCT NeuroComp and the Funcap for support. FL acknowledges support  from the European Union's Horizon research and innovation program under the Marie Sklodowska-Curie Grant Agreement No. 101066790 and from the program TAlent in ReSearch@University of Padua -- STARS@UNIPD  (project  BRAINCIP---Brain criticality and information processing). This research was supported by the Intramural Research Program of the National Institutes of Mental Health (NIMH), USA, ZIAMH002797, ZIAMH002971.

\end{acknowledgments}

\bibliography{references}


\onecolumngrid 

\clearpage 

\appendix

\centering{\Large {\bf Supplementary Materials}}

\vspace{4cm} 


\renewcommand{\thesection}{S\arabic{section}} 

\renewcommand{\thesubsection}{S\arabic{section}.\arabic{subsection}}
\renewcommand{\theequation}{S\arabic{equation}}
\renewcommand{\thefigure}{S\arabic{figure}}
\renewcommand{\thetable}{S\arabic{table}}  

\setcounter{figure}{0}

\clearpage

\section*{Supplementary Figures}

\vspace{2cm}

\begin{figure}[!h]                              
    \centering
    \includegraphics[width=0.87\linewidth]{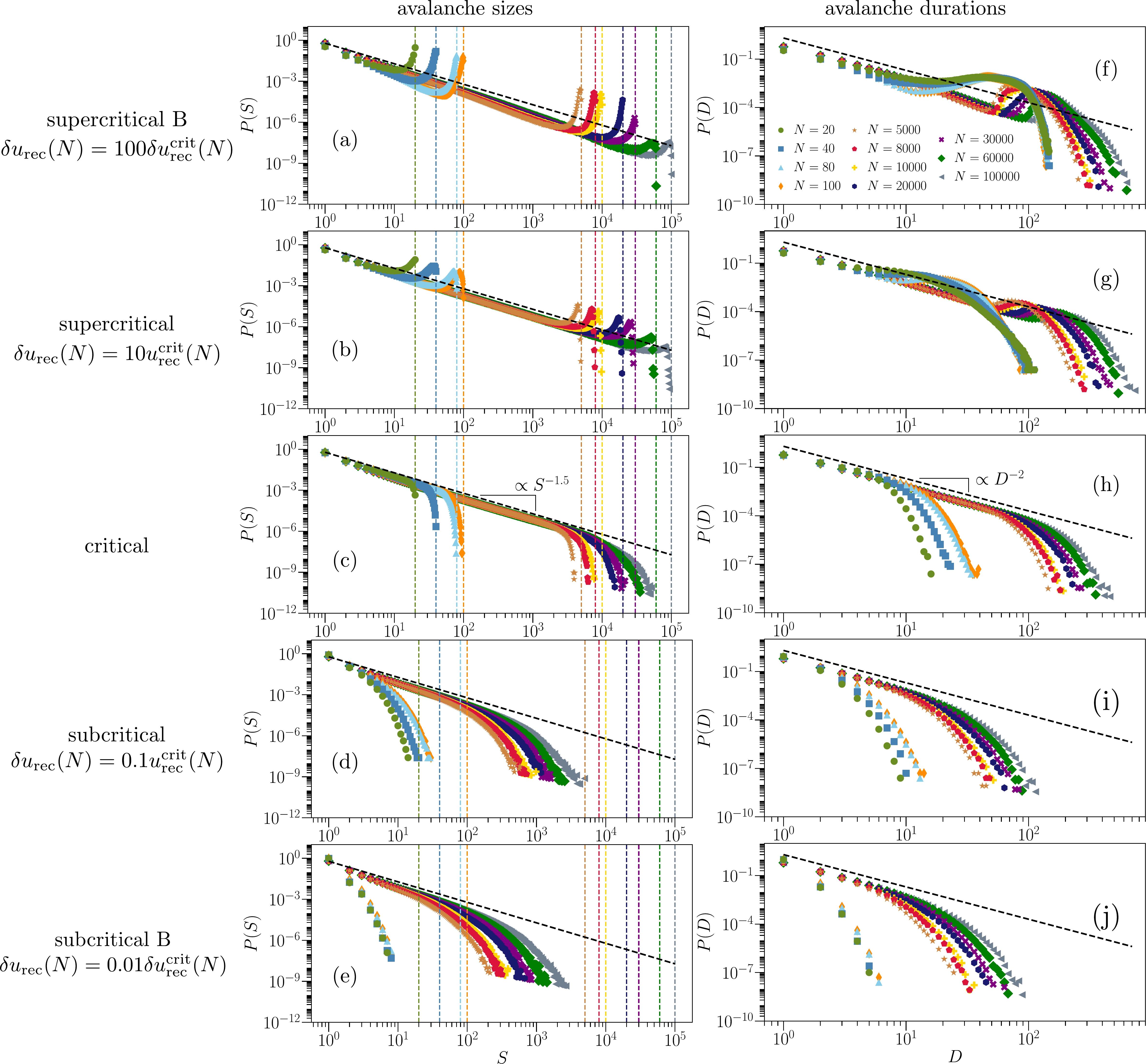}      
    \caption{Distributions of avalanche size $P(S)$ (a-e) and duration $P(D)$ (f-j) for the IF model with $\pin = 20 \%$ inhibitory neurons in the subcritical (bottom and second-to-last row), critical (middle row) and supercritical (second and top row) state, for system sizes $ N \in [2 \cdot 10^1,10^5]$. For the subcritical and supercritical data, we set $\durec(N) = 0.1 \durecCrit(N)$ (subcritical) or $\durec(N) = 0.01 \durecCrit(N)$ (subcritical B), and $\durec(N) = 10 \durecCrit(N)$ (supercritical) or $\durec(N) = 100 \durecCrit(N)$ (supercritical B), respectively, where $\durecCrit(N)$ is the value that sets the IF network of size $N$ to the critical state, reported in Table (\ref{tableS:durec}). In (a-e), the vertical dashed lines indicate the system size of the corresponding color-matching curve of $P(S)$. Results are averaged over $2000$ IF network configurations, and over $2 \cdot 10^{4}$ avalanches for each configuration. }       
    \label{figS:aval-dists}     
\end{figure}        

\clearpage

 \begin{figure}
    \centering
    \includegraphics[width=\linewidth]{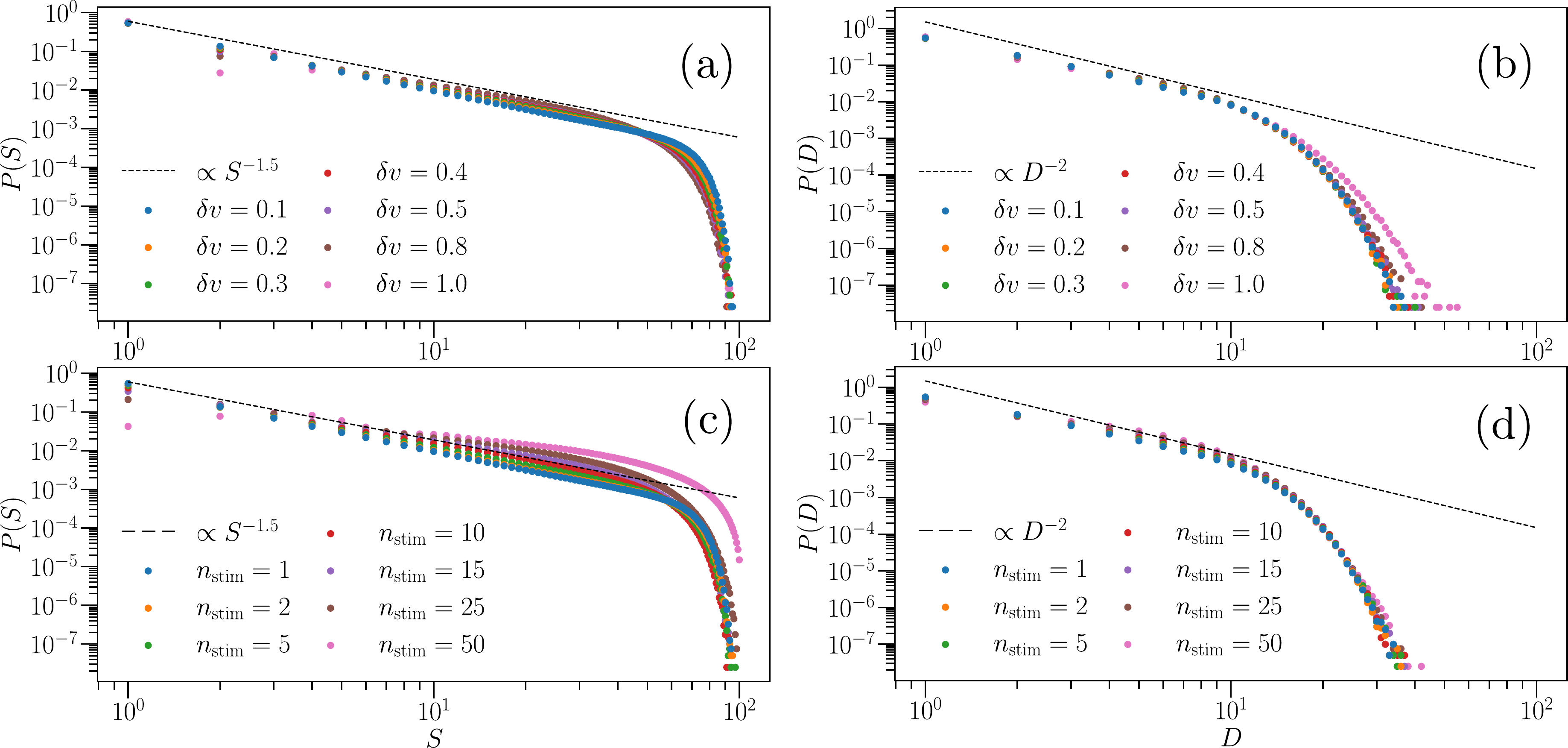}   
    \caption{Distributions of avalanche size $P(S)$ (left) and avalanche duration $P(D)$ (right) measured in IF networks with $N=100$ neurons as a function of the external input $\delta v$ (a-b) or the number $n_\mathrm{stim}$ of neurons that are randomly selected  and  receive an external input $\delta v = 0.1$ at each timestep (c-d). For all simulations, we set $\durec = \durecCrit(N=100)=1.26\cdot10^{-3}$. Results are averaged over 2000 IF network configurations and $2 \cdot 10^4$ avalanches per configuration.}      
    \label{figS:aval-dists-vs-dv-and-nstim} 
\end{figure}      
 
\begin{figure}    
    \centering  
    \includegraphics[width=0.5\linewidth]{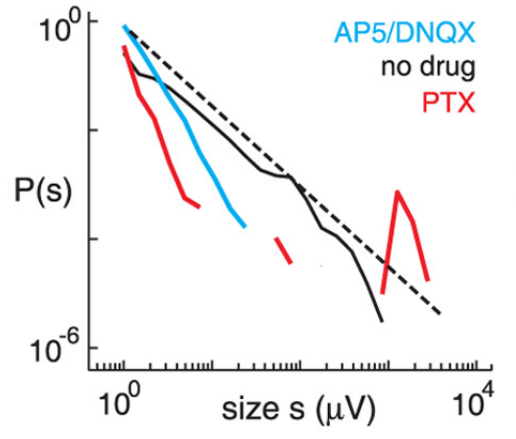}  
    \caption{Distributions of avalanche size for normal (no-drug, black), disinhibited (PTX, red), and hypoexcitable (AP5/DNQX, blue) cortical cultures. Avalanche size $S$ is defined as the absolute sum of all negative local field potential (nLFP) amplitudes within a spatiotemporal cluster of neuronal activity. The dashed line is the power law $P(S) \propto S^{-1.5}$. Adapted from \cite{shew_neuronal_2009}. Copyright 2009 Society for Neuroscience. 
    }   
    \label{figS:aval-dists-exp}  
\end{figure}

\begin{figure} 
    \centering
    \includegraphics[width=\linewidth]{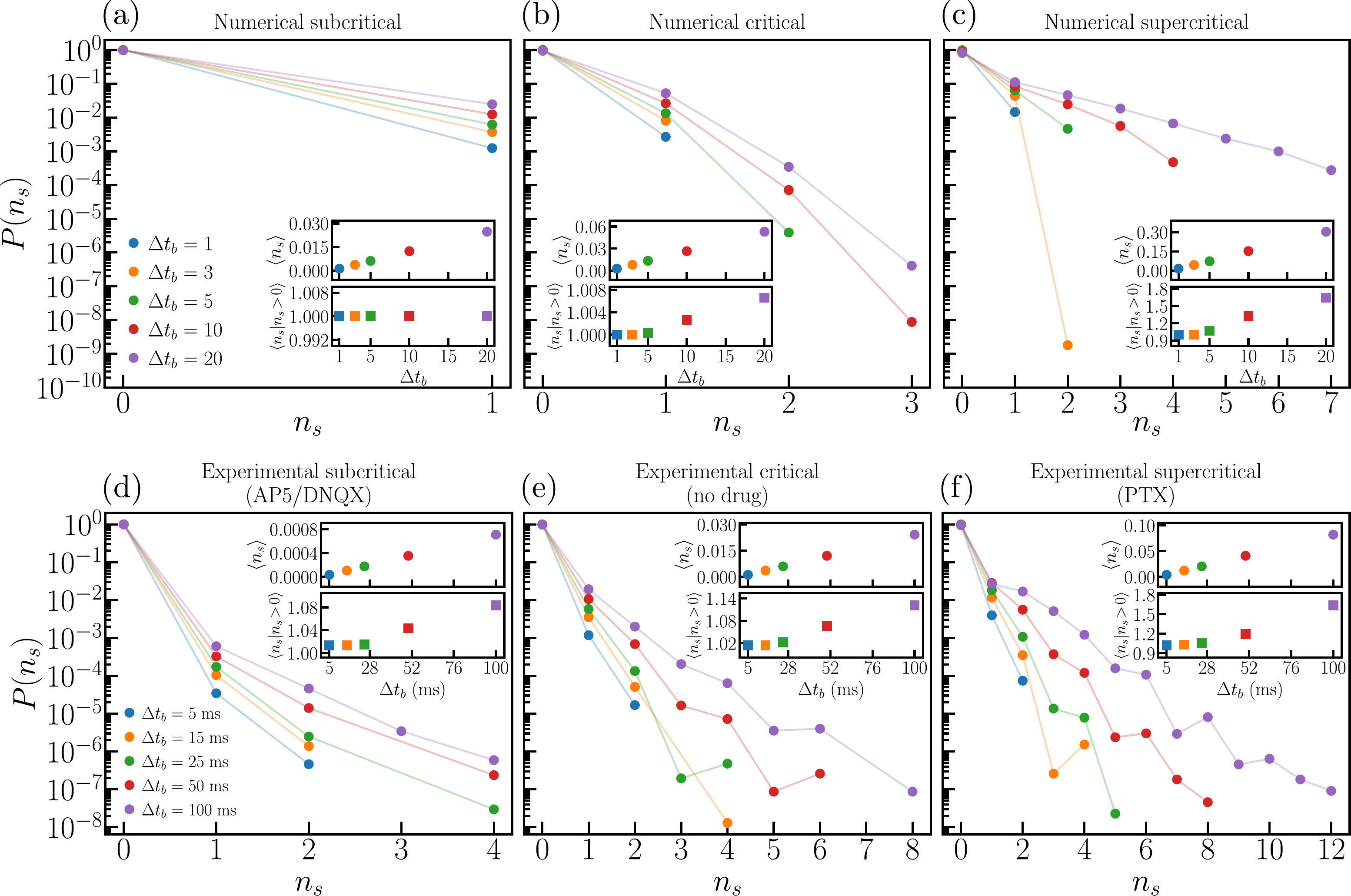} 
    \caption{Distributions of the number $n_s$ of spikes per timebin for several bin sizes $\Delta t_b$ measured from IF networks ($N=100$ neurons; 5 different IF network configurations per distribution; $N_b = 10^7$ timebins per configuration) in the subcritical (a), critical (b) and supercritical (c) state, as well as from experimental data obtained from cortex slice cultures  treated with a combination of AP5/DNQX (d), no-drug cultures (e) and cultures treated with PTX (f), whose neuronal dynamics were respectively classified as subcritical, critical and supercritical in \cite{shew_neuronal_2009} ($N = 60$ electrodes; 4 (d) or 5 (e, f) independent 1 hour recordings per distribution). Insets: average number of spikes per bin, $\langle n_s \rangle$ (top). Conditional average, $\langle n_s | n_s > 0\rangle$ (bottom), taking only non-empty timebins, as a function of $\Delta t_b$.} 
    \label{figS:spike-per-bin-dists} 
\end{figure}   

\begin{figure} 
    \centering
    \includegraphics[width=\linewidth]{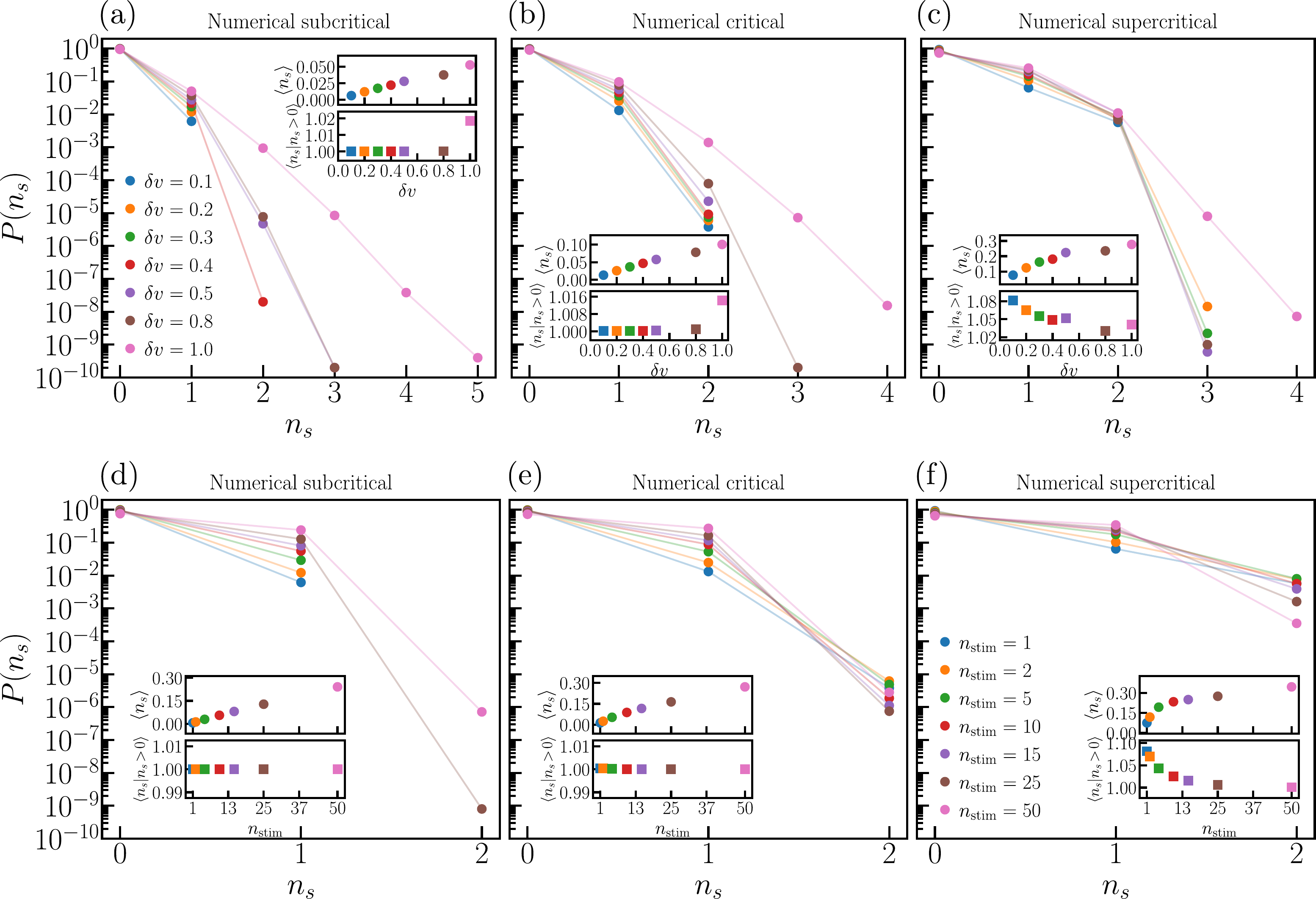}  
    \caption{Same as in Fig.\ \ref{figS:spike-per-bin-dists} but varying the external input $\delta v$ (a-c) or the number $n_{\mathrm{stim}}$ of simultaneously stimulated neurons (d-f) in IF networks.} 
    \label{figS:spike-per-bin-vs-dv-and-n_stim}  
\end{figure}

\begin{figure}
    \centering
    \includegraphics[width=\linewidth]{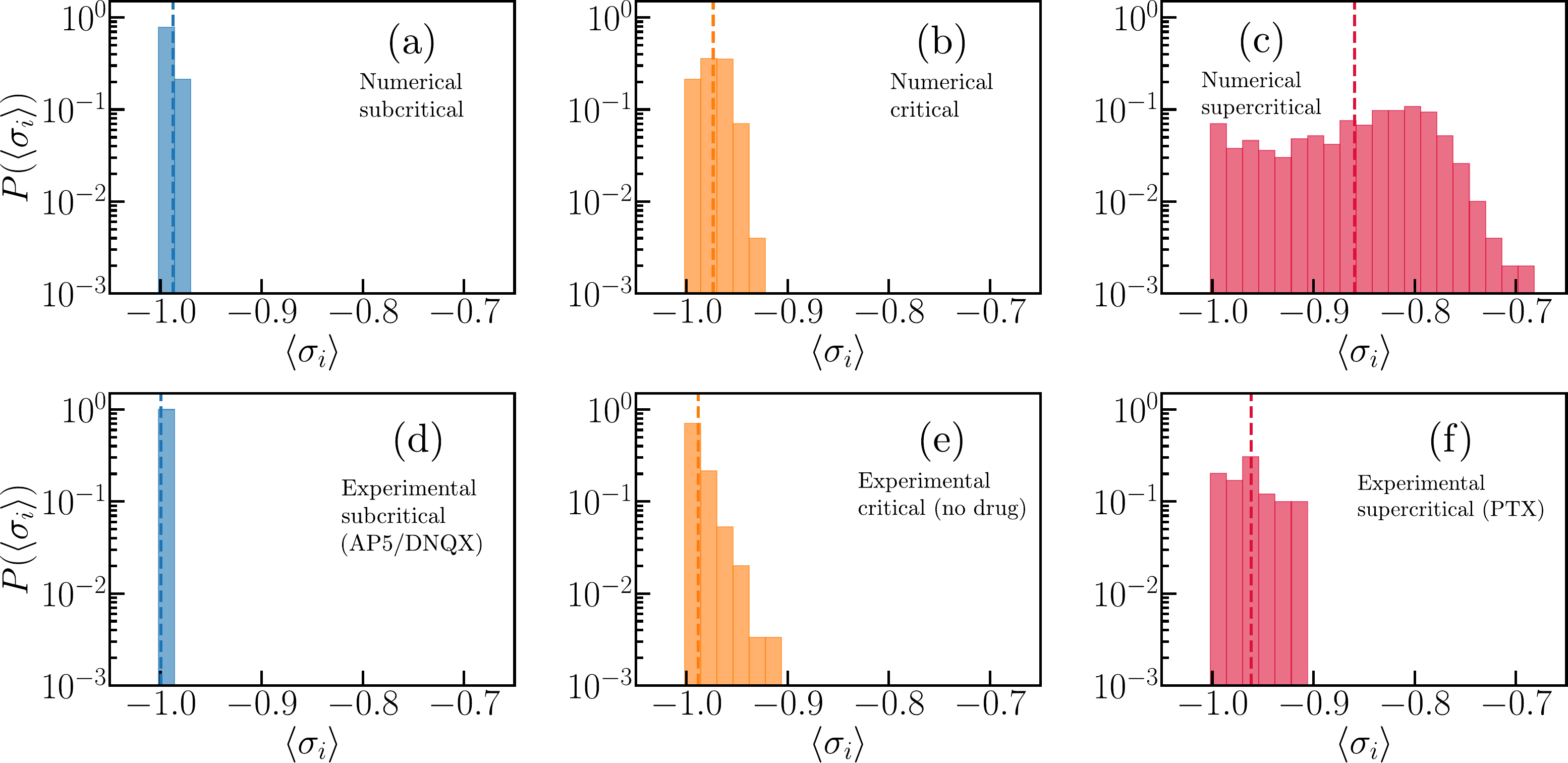} 
    \caption{Distributions of the average local activity $\aveSigmaData$ measured from IF networks ($N=100$ neurons; 5 different IF network configurations per distribution; $N_b = 10^7$ timebins per configuration) in the subcritical (a), critical (b) and supercritical (c) state, as well as from experimental data obtained from cortex slice cultures  treated with a combination of AP5/DNQX (d), no-drug cultures (e) and cultures treated with PTX (f), whose neuronal dynamics were respectively classified as subcritical, critical and supercritical in \cite{shew_neuronal_2009} ($N = 60$ electrodes; 4 (d) or 5 (e,f) independent 1 hour recordings per distribution). The vertical dashed lines indicate the mean of the distributions.} 
    \label{figS:aveSigma-dists}
\end{figure} 

 \begin{figure}
    \centering
    \includegraphics[width=\linewidth]{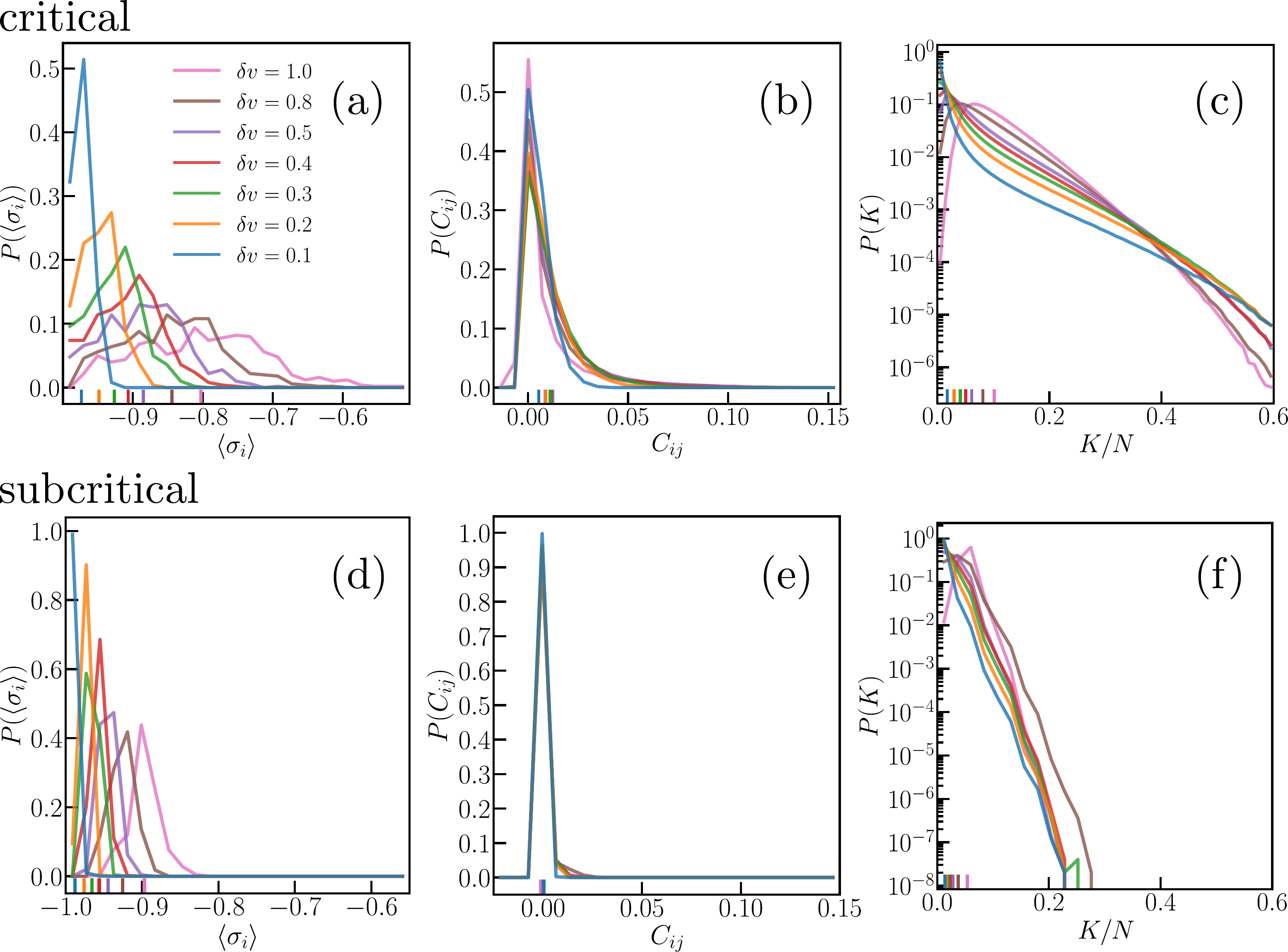}  
    \caption{Distributions of $\aveSigmaData$ (a), $\CijData$ (b), and $\PKData$ (c), measured in IF networks with $N=100$ neurons in the critical state, as a function of the external input $\delta v$ applied to a randomly selected neuron at each timestep. The colored bars on the x-axis indicate the mean of the corresponding color-coded distributions. 
    (d-f) Same as in (a-c) but for IF networks in the subcritical state.   
    Results are averaged over 5 IF network configurations and evaluated over $N_b = 10^7$ time bins per configuration.}    
    \label{figS:dists-vs-dv} 
\end{figure}   
 
 \begin{figure}
    \centering
    \includegraphics[width=\linewidth]{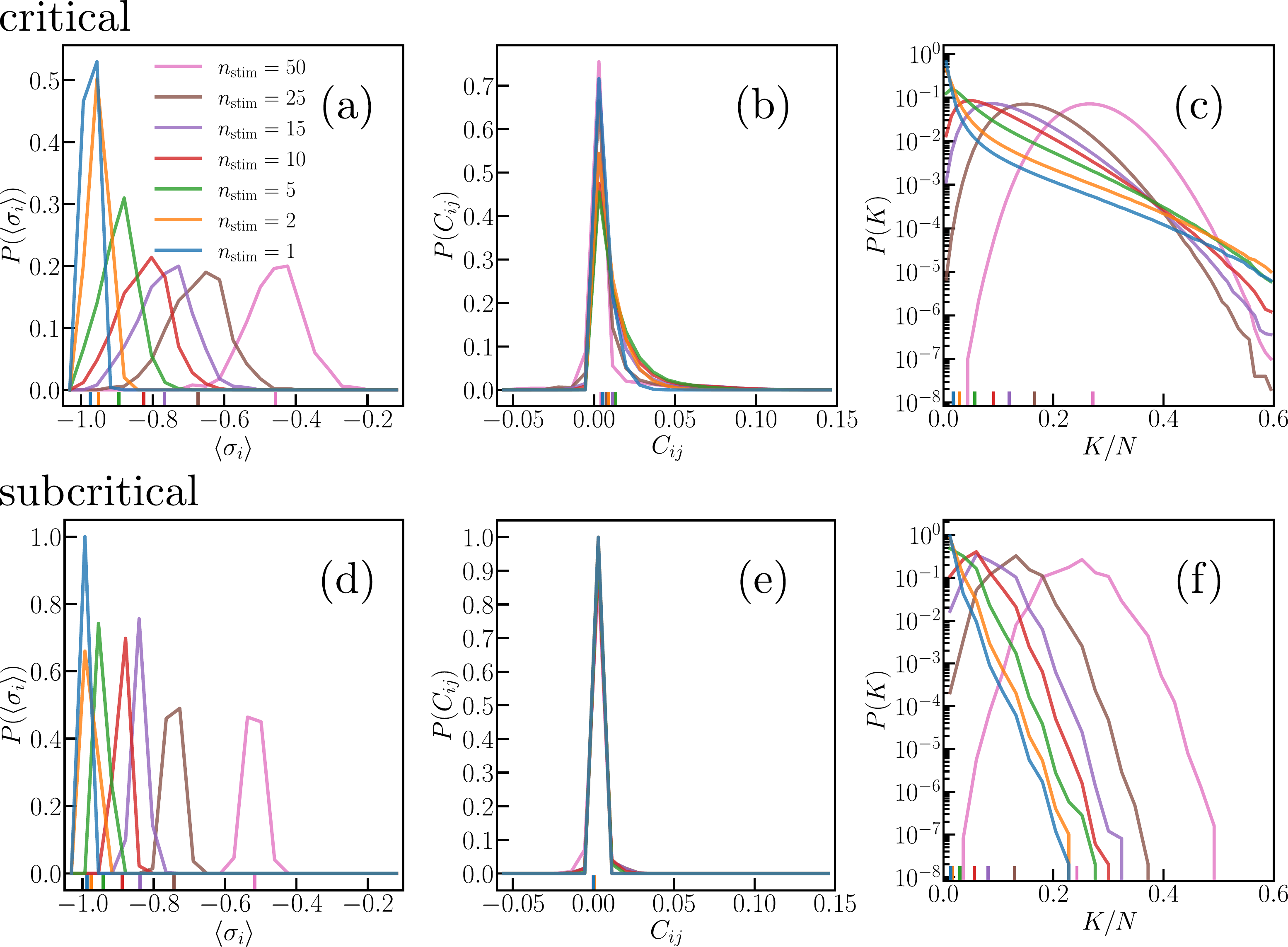}   
    \caption{Same as in Fig.\ \ref{figS:dists-vs-dv} but instead as a function of the number $n_\mathrm{stim}$ of randomly selected neurons  that simultaneously receive an external input $\delta v = 0.1$ at each timestep.}     
    \label{figS:dists-vs-n_stim} 
\end{figure}

\begin{figure}
    \centering
    \includegraphics[width=\linewidth]{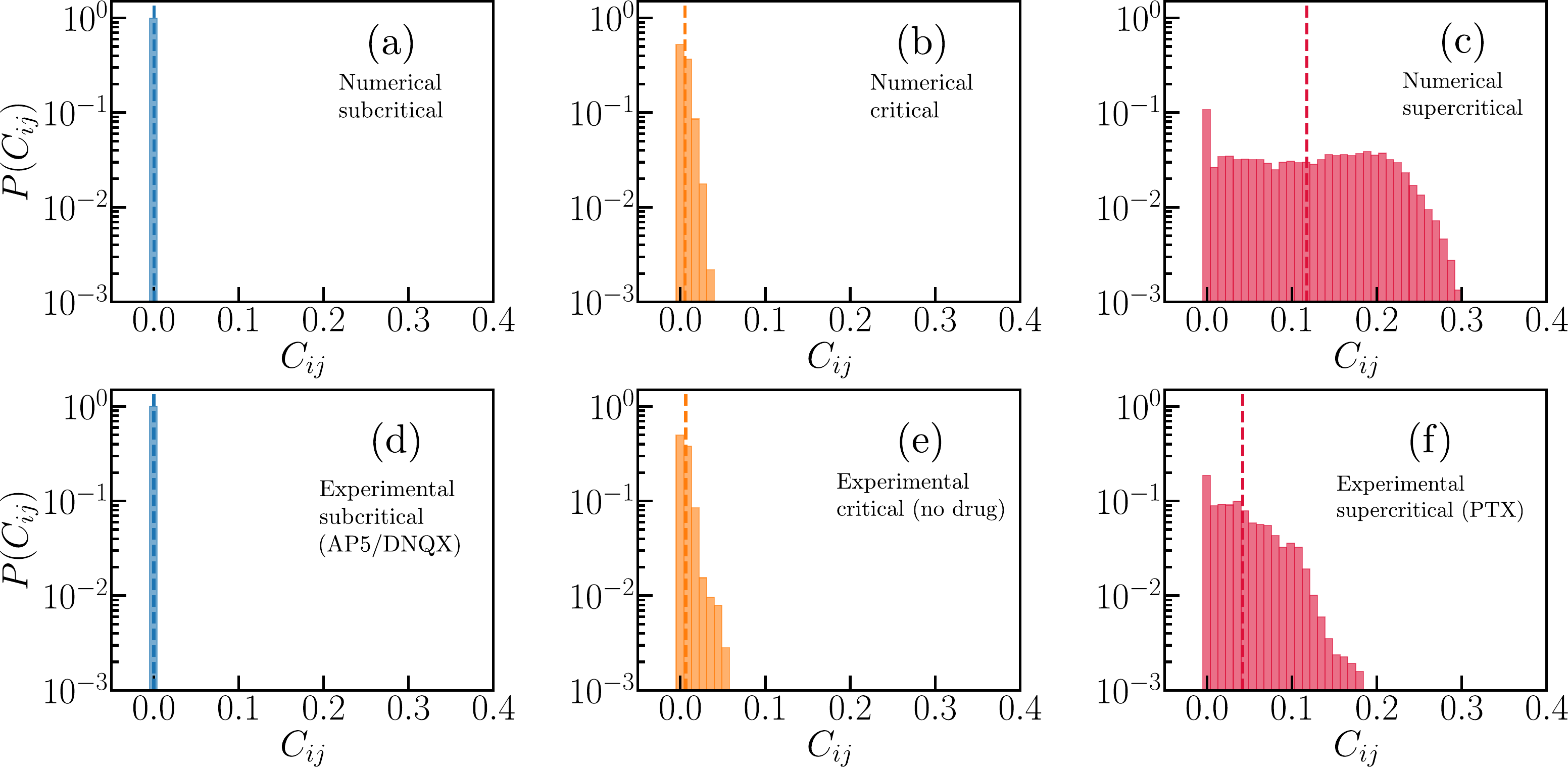}  
    \caption{Distributions of the correlation functions $C_{ij}$ measured from IF networks ($N=100$ neurons; 5 different IF network configurations per distribution; $N_b = 10^7$ timebins per configuration) in the subcritical (a), critical (b) and supercritical (c) state, as well as from experimental data obtained from cortex slice cultures  treated with a combination of AP5/DNQX (d), no-drug cultures (e) and cultures treated with PTX (f), whose neuronal dynamics were respectively classified as subcritical, critical and supercritical in \cite{shew_neuronal_2009} ($N = 60$ electrodes; 4 (d) or 5 (e,f) independent 1 hour recordings per distribution). The vertical dashed lines indicate the mean of the distributions.} 
    \label{figS:Cij-dists} 
\end{figure}

\begin{figure}          
    \centering
    \includegraphics[width=0.8\linewidth]{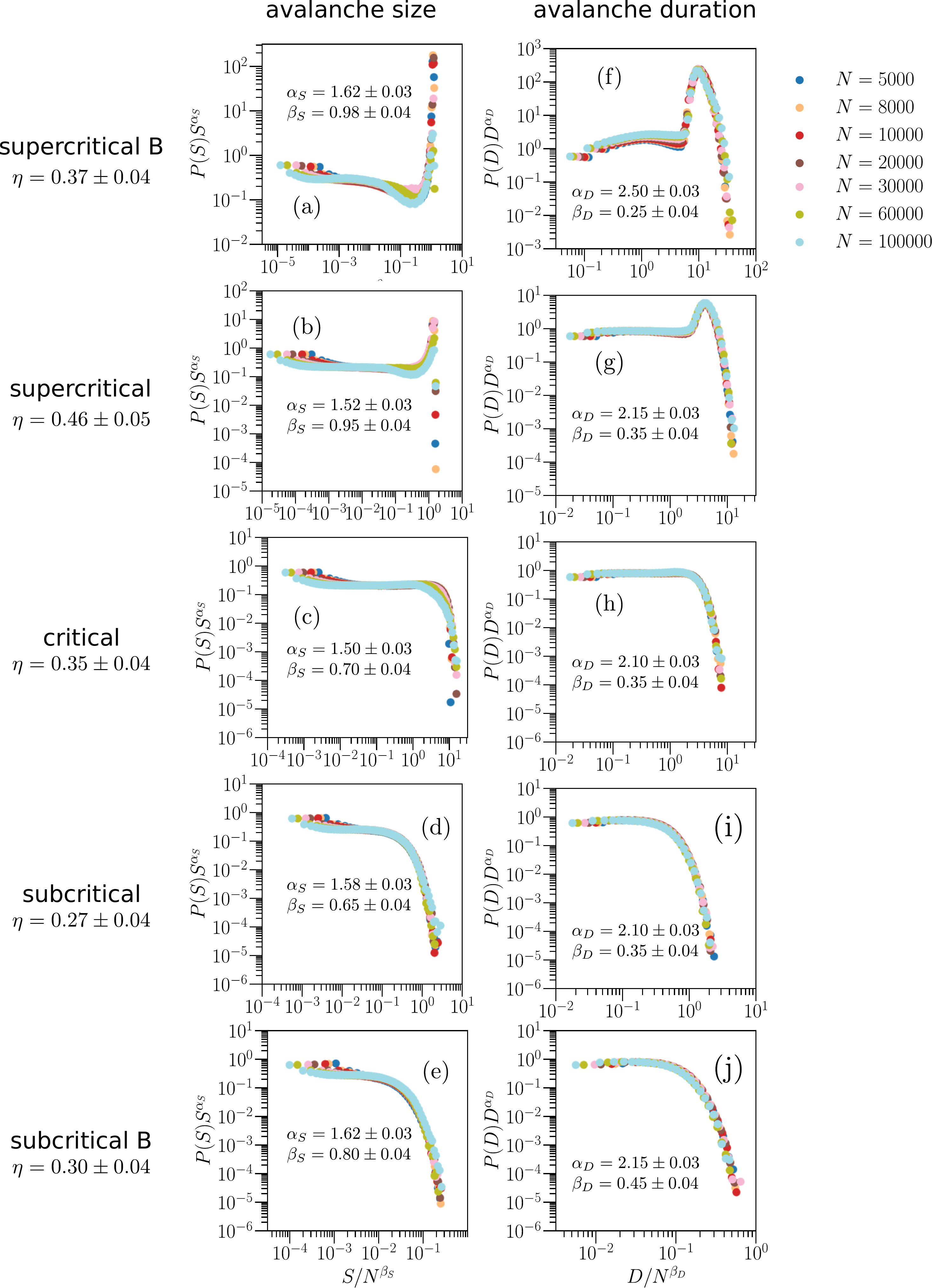} 
    \caption{Collapse of the numerical distributions of avalanche sizes $S$ (a-e) and durations $D$ (f-j), for different sizes $N \in [5 \cdot 10^3, 10^5]$, onto a universal curve $\tilde{P}(S) = S^{\alpha_S} \mathcal{F}(S/N^{\beta_S})$ and $\tilde{P}(D) = D^{\alpha_D} \mathcal{F}(D/N^{\beta_D})$, respectively, considering IF networks below, at and above the critical state. $\alpha_S$ is the exponent that characterizes the intermediate power-law regime of $P(S)$ while $\beta_S$ is the exponent of the scaling of the cut-off, and analogously for $\alpha_D$ and $\beta_D$ for $P(D)$. Since $\alpha_D > 2$ for all cases,
    the allometric exponent is expected to depend only on $\alpha_S$ and $\beta_S$, according to $ \eta = \beta_S \cdot ( 2 - \alpha_S ) $ (see \cite{simoes_allometric_2026}). The analytical prediction for $\eta$ is presented in the leftmost part of the figure for each case: subcritical B ($\durec(N) = 0.001 \durecCrit(N)$), subcritical ($\durec(N) = 0.01 \durecCrit(N)$), critical ($\durec(N) = \durecCrit(N)$, with the values of $\durecCrit(N)$ for each $N$ reported in Table \ref{tableS:durec}), supercritical ($\durec(N) = 10 \durecCrit(N)$) and supercritical B ($\durec(N) = 100 \durecCrit(N)$). }
    \label{figS:avalanche-dists-collapse}
\end{figure}           

\clearpage  

\begin{figure} 
    \centering 
    \includegraphics[width=\linewidth]{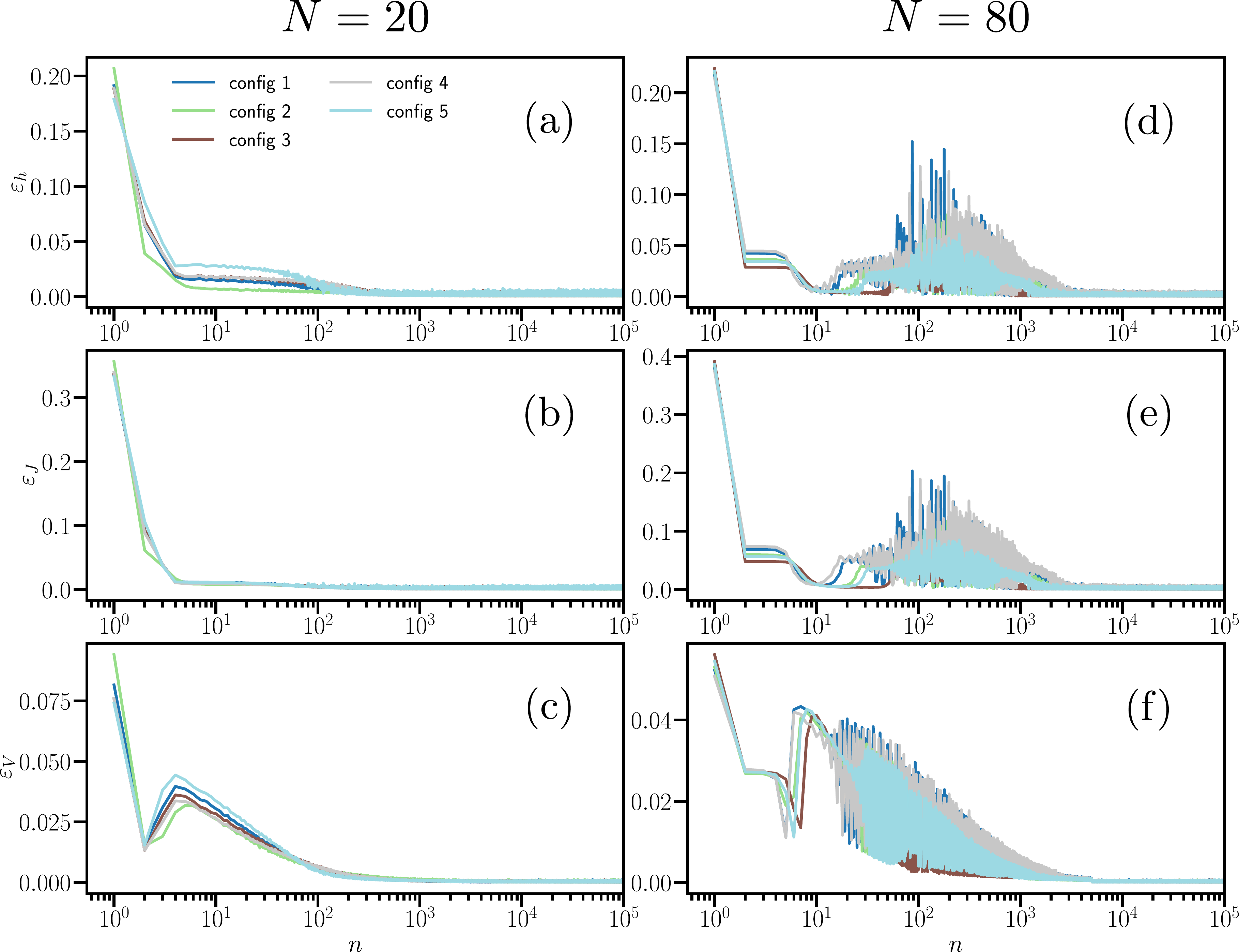} 
    \caption{Error functions $\varepsilon_h = \sqrt{ \frac{1}{N} \sum_{i} \left( \aveSigmaMEM - \aveSigmaData \right)^2}$ (a, d), $\varepsilon_J = \sqrt{ \frac{2}{N \cdot (N - 1)} \sum_{i<j} \left( \aveTwoMEM - \aveTwoData \right)^2}$ (b, e), and $\varepsilon_V = \sqrt{ \frac{1}{N_K} \sum_{K:\ \PKData>10^{-5}} \left( \PKMEM - \PKData \right)^2}$ (c, f), plotted as a function of the number of Boltzmann Machine iterations $n$, for critical IF networks of size $N=20$ (a-c) and $N=80$ (d-f), for five network configurations per size. The sum in $\varepsilon_V$ includes only $K$ values with $\PKData>10^{-5}$, in accordance with the threshold explained in the main text used for fitting $V_K$ from the numerical data, and $N_K$ is the number of such $K$ values.}    
    \label{figS:RMSD}  
\end{figure}

\clearpage

 \begin{figure}                
    \centering
    \includegraphics[width=0.9\linewidth]{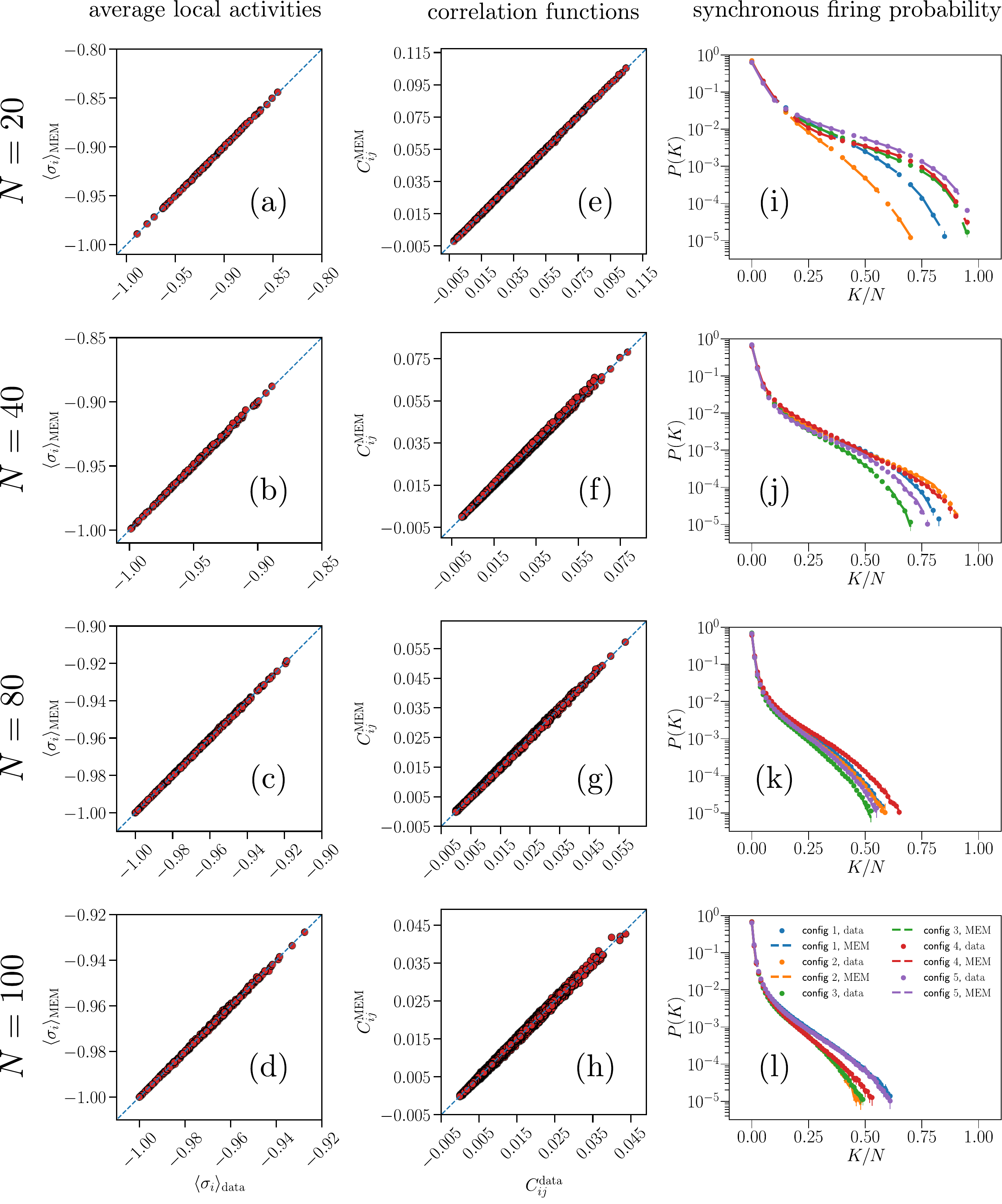}      
    \caption{Comparison of the average local activities (a-d), correlation functions (e-h) and synchronous firing probability (i-l) from 5 IF networks of size $N \in \{ 20 , 40 , 80 , 100 \} $ in the critical state (data) to the analogous quantities in the associated maximum entropy model (MEM). In (a-h), the blue dashed lines are the bisector $y = x$. For the IF model data, results are averaged over $N_b = 10^7$ timebins. For the maximum entropy models, results are averaged over $M_c = 3 \cdot 10^6$ spin configurations, and over 100 random initial spin configurations. Error bars are given by the standard error, and are overall smaller or equal to the symbol size.}      
    \label{figS:quality-of-fit-critical}     
\end{figure}

\begin{figure}               
    \centering
    \includegraphics[width=0.9\linewidth]{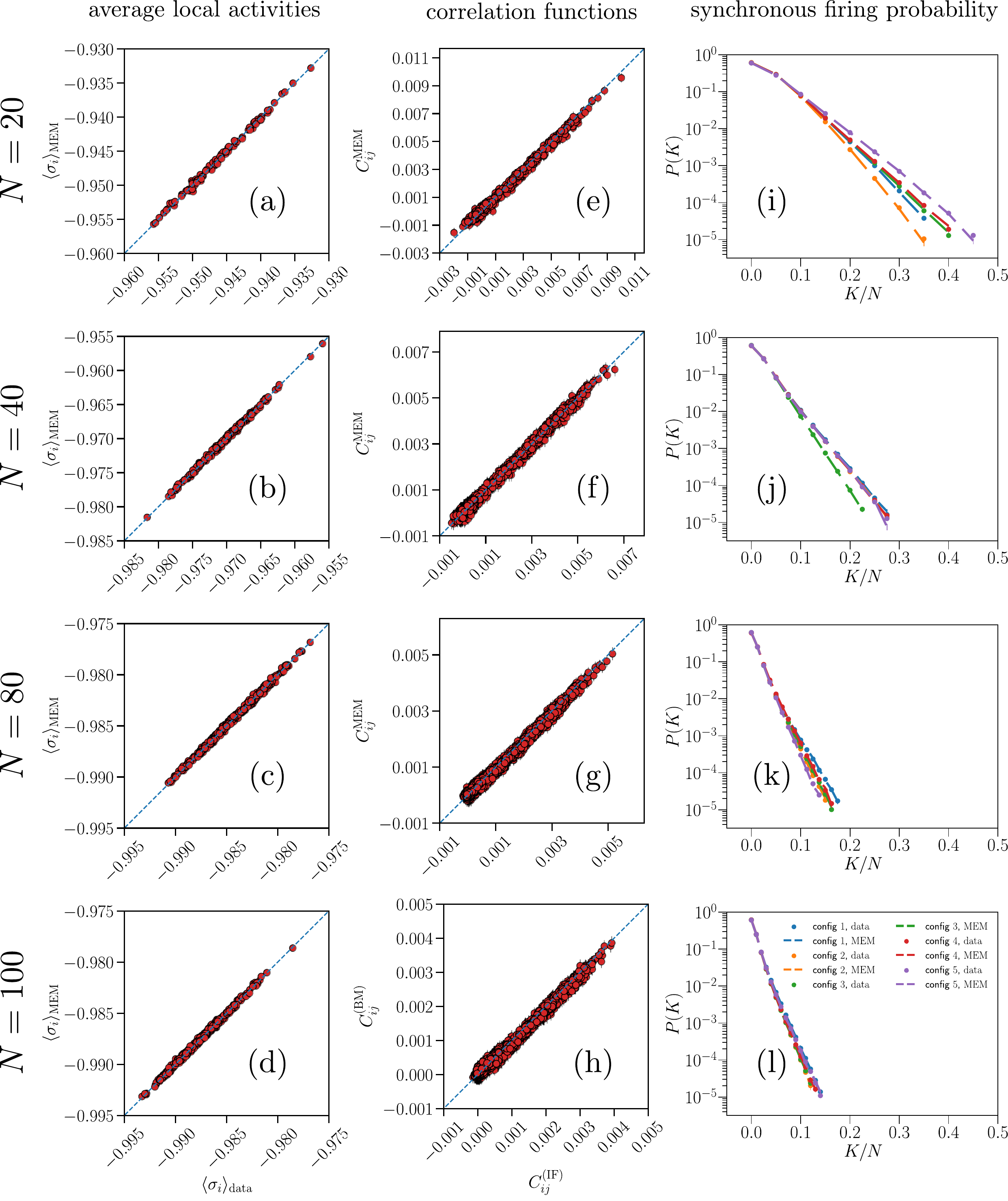}      
    \caption{Same as in Fig. (\ref{figS:quality-of-fit-critical}) for IF networks in the subcritical state.}      
    \label{figS:quality-of-fit-subcritical}     
\end{figure}

\begin{figure}                
    \centering
    \includegraphics[width=0.9\linewidth]{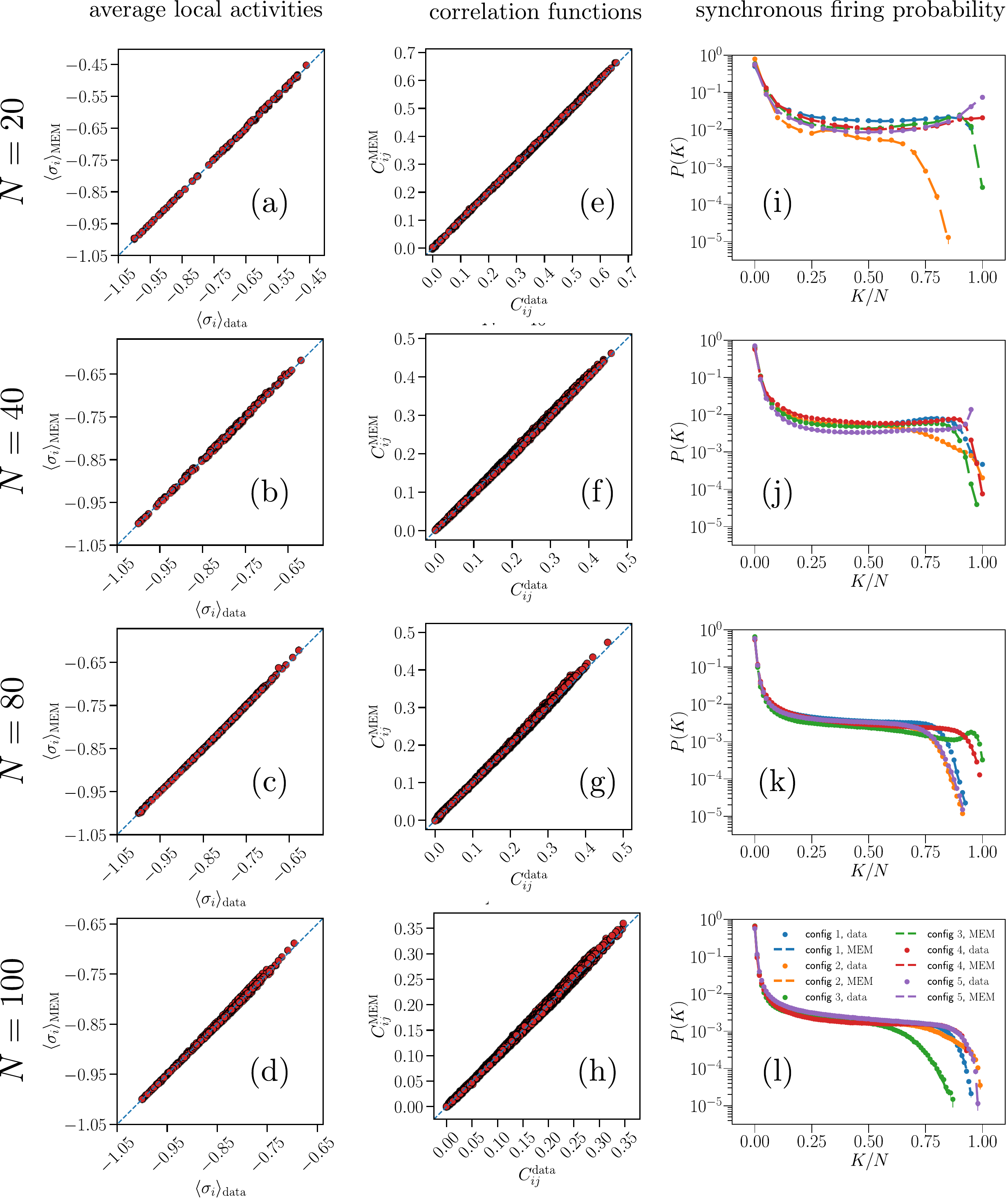}      
    \caption{Same as in Fig. (\ref{figS:quality-of-fit-critical}) for IF networks in the supercritical state.}      
    \label{figS:quality-of-fit-supercritical}     
\end{figure}             
 
\begin{figure}                 
    \centering
    \includegraphics[width=0.9\linewidth]{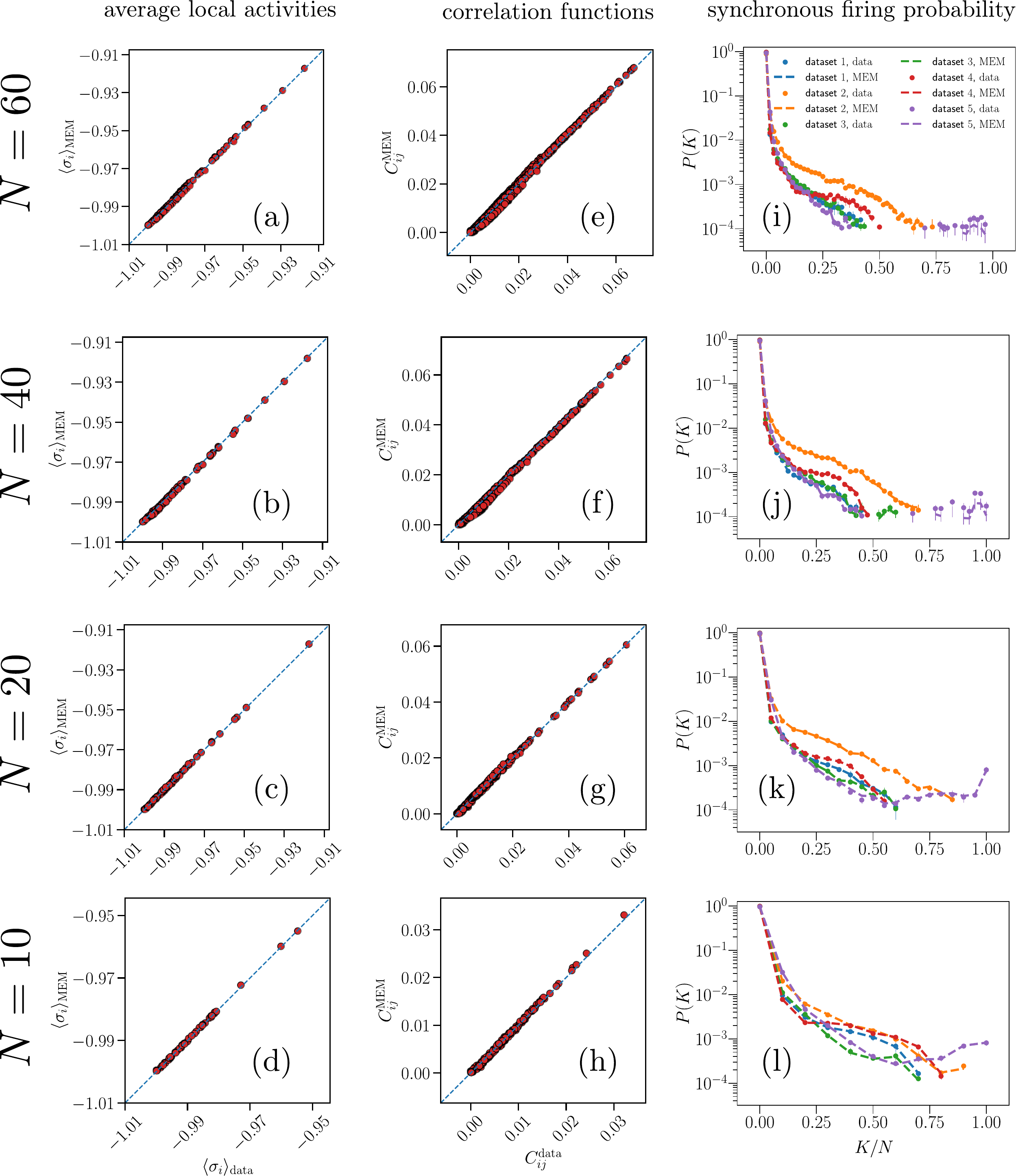}      
    \caption{Same as in Figs.\ (\ref{figS:quality-of-fit-critical}-\ref{figS:quality-of-fit-supercritical}) for 5 experimental recordings of $N=60$ electrodes, as well as random subsamplings at sizes $N \in \{10,20,40\}$, using the same cortical culture, whose dynamics were identified as critical in \cite{shew_neuronal_2009}. For the experimental data, results are averaged over $N_b \approx 1.4 \cdot 10^5$ timebins.}       
    \label{figS:quality-of-fit-experimental-critical}      
\end{figure}              

 \begin{figure}                 
    \centering
    \includegraphics[width=0.9\linewidth]{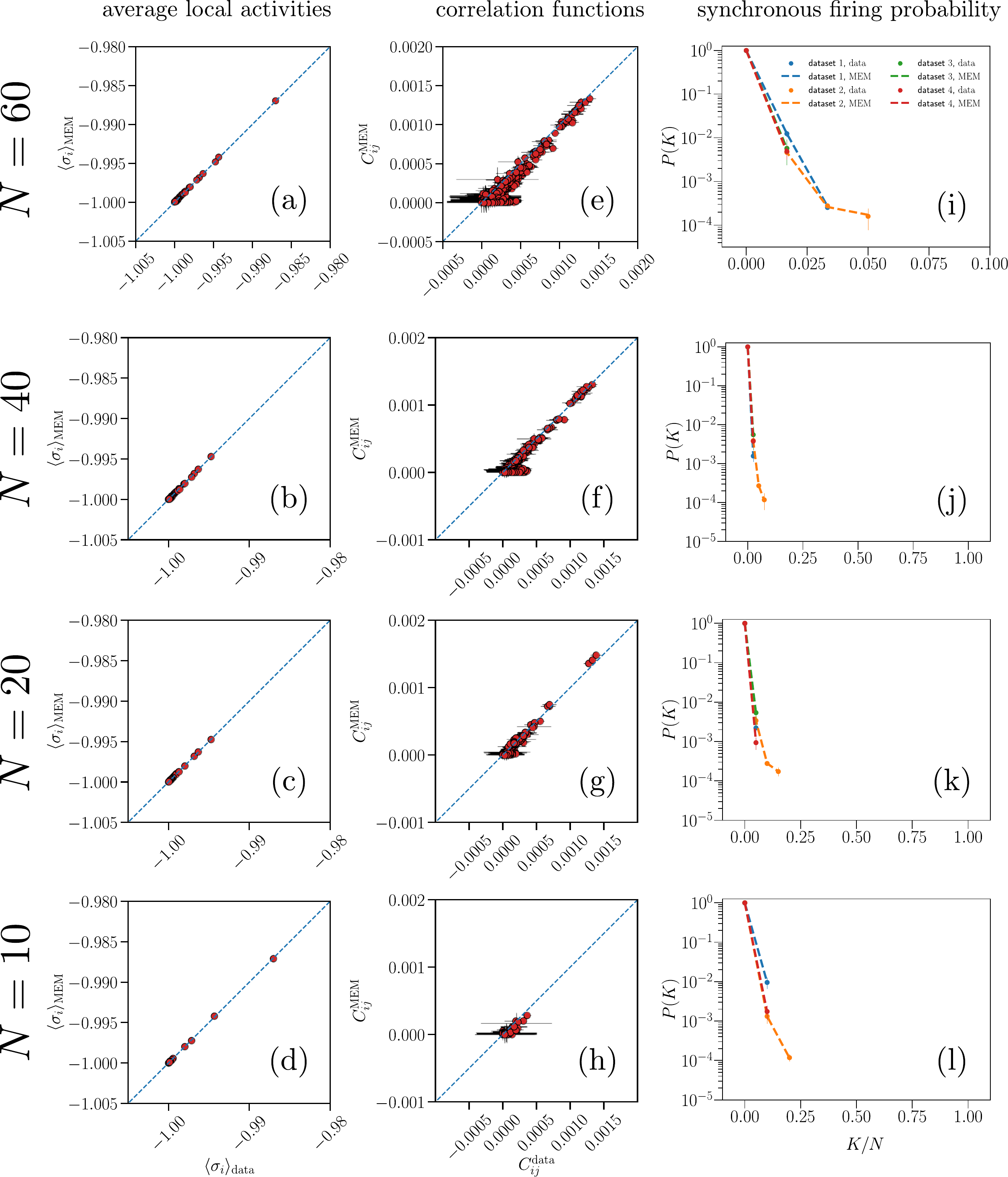}        
    \caption{Same as in Fig.\ (\ref{figS:quality-of-fit-experimental-critical}) for 4 experimental recordings of $N=60$ electrodes and random subsamplings at sizes $N \in \{10,20,40\}$, using the same cortical culture treated with a mixture of AP5/DNQX drugs, whose dynamics were indentified as subcritical in \cite{shew_neuronal_2009}.}        
    \label{figS:quality-of-fit-experimental-subcritical}      
\end{figure}        
      
 \begin{figure}                
    \centering
    \includegraphics[width=0.9\linewidth]{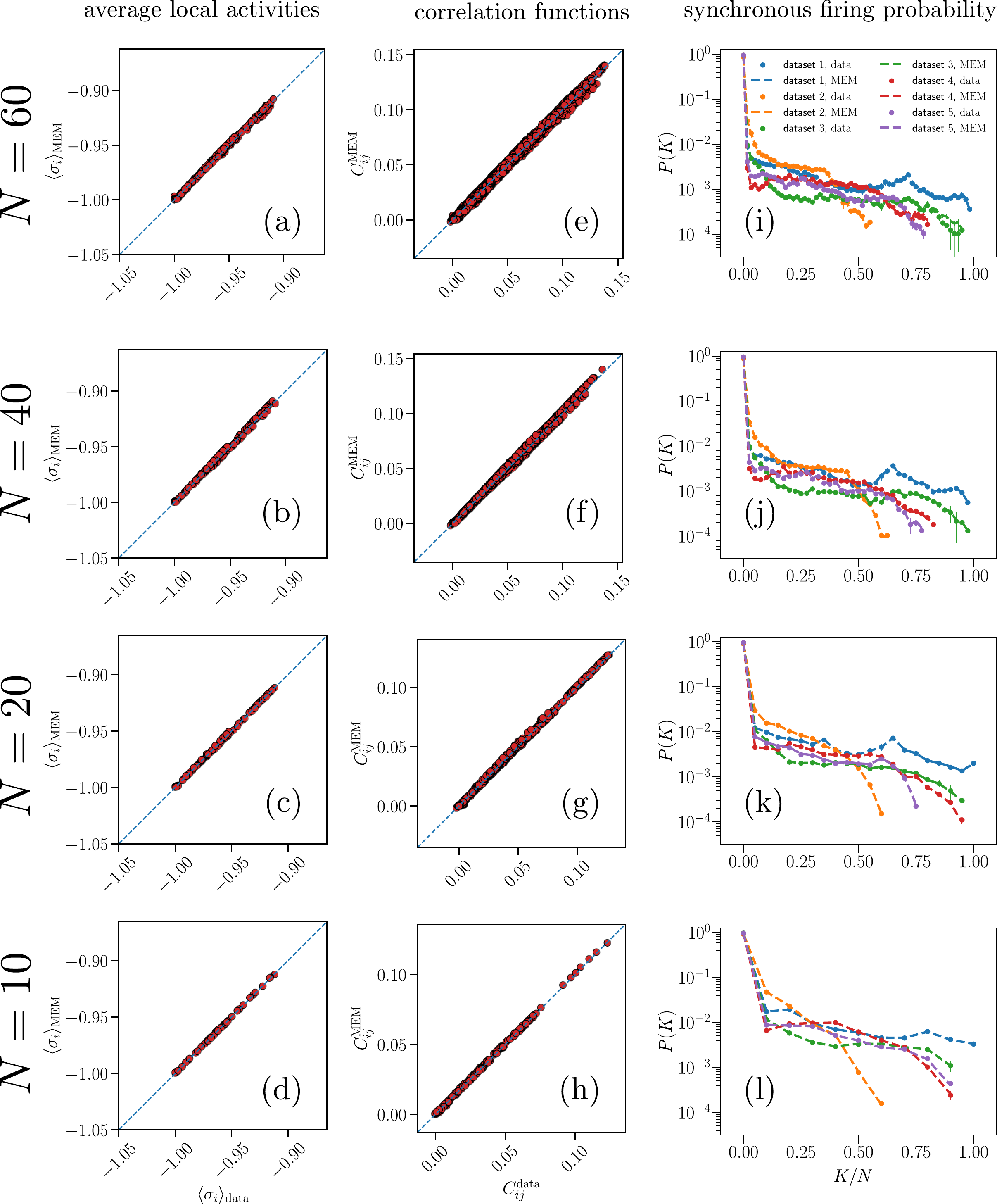}    
    \caption{Same as in Fig.\ (\ref{figS:quality-of-fit-experimental-critical}) for 5 experimental recordings of $N=60$ electrodes and random subsamplings at sizes $N \in \{10,20,40\}$, using the same cortical culture treated with PTX drug, whose dynamics were identified as supercritical in \cite{shew_neuronal_2009}.}       
    \label{figS:quality-of-fit-experimental-supercritical}     
\end{figure}             
   
\begin{figure}                 
    \centering
    \includegraphics[width=\linewidth]{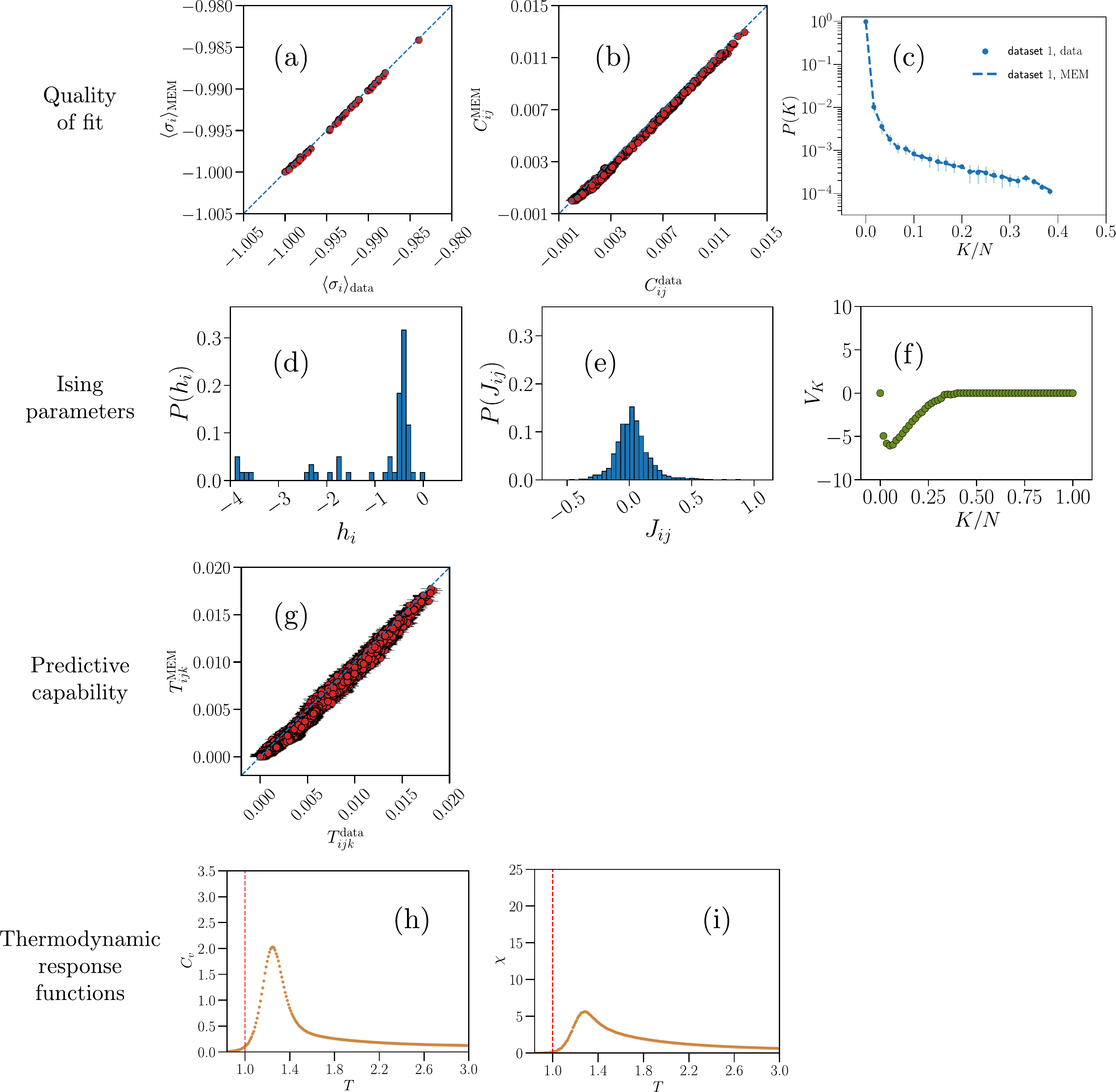}       
    \caption{ME analysis of an experimental recording of $N=60$ electrodes from a cortical culture treated with AP5 drug, whose dynamics were identified as subcritical in \cite{shew_neuronal_2009}. (a-c) Comparison of the (a) average local activities, (b) correlation functions and (c) synchronous firing probability from the experimental data to the analogous quantities in the associated ME model. (d-f) Parameters of the ME model: (a) Distribution of fields $h_i$, (e) distribution of interaction constants $J_{ij}$ and (f) plots of the potentials $V_K$ as a function of $K/N$. (g) Comparison of the three-point correlation functions $T_{ijk}$ between the experimental data and the prediction of the ME model. (h-i) Thermodynamic response functions of the ME model as a function of the temperature $T$: (h) Specific heat $C_v$ and (i) susceptibility $\chi$.
    }       
    \label{figS:AP5-results}      
\end{figure}

 \begin{figure}               
    \centering
    \includegraphics[width=\linewidth]{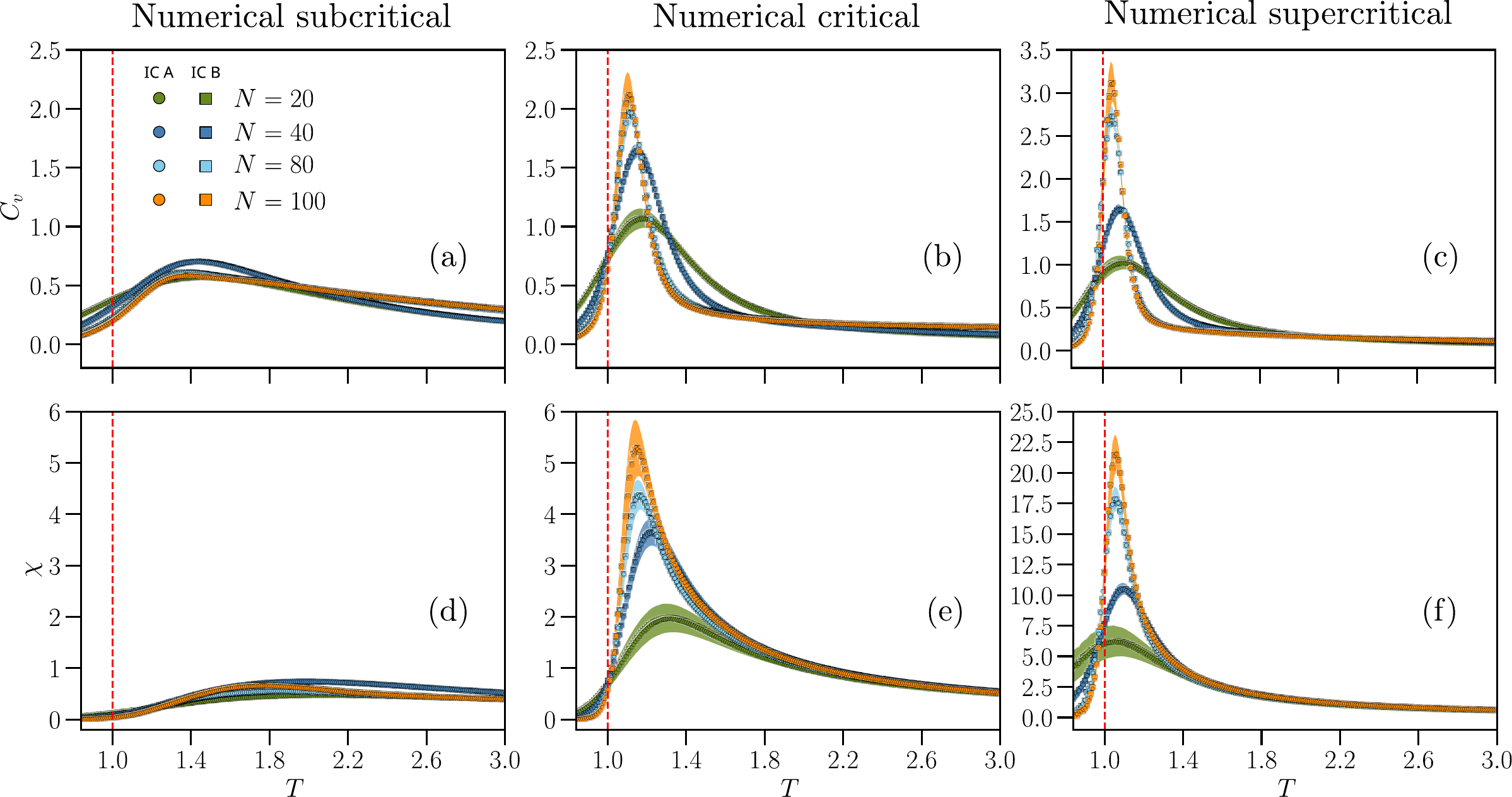}       
    \caption{Specific heat $C_v$ (a–c) and susceptibility $\chi$ (d–f) as functions of temperature $T$ for K-pairwise Ising models fitted to IF networks of size $N \in \{20, 40, 80, 100\}$ in subcritical (left column), critical (center column), and supercritical states (right column), considering two different initial conditions for the Monte Carlo simulations: Initial condition A - starting with random $\sigma_i$ (circle symbols, same results as those presented in Fig.\ (\ref{fig:thermodynamics}) of the main text); Initial condition B - starting with all $\sigma_i = -1$ (square symbols).   
    Vertical dashed lines indicate $T = 1$, where the $T$-parameterized probability (Eq.\ (\ref{eq:PTemp}) in the main text) matches the maximum entropy distribution (Eq.\ (\ref{eq:maxEntProbGeneral}) in the main text) that fits the data. For each $T$, results are averaged over $M_c = 3 \cdot 10^6$ spin configurations, and $100$ random initial spin configurations. The colored shaded areas around the curves of $C_v$ and $\chi$ represent the standard error obtained from $5$ different IF network configurations for each $N$.    
    }      
    \label{figS:thermodynamics-IC-comparison}       
\end{figure}               

 \begin{figure}                
    \centering
    \includegraphics[width=\linewidth]{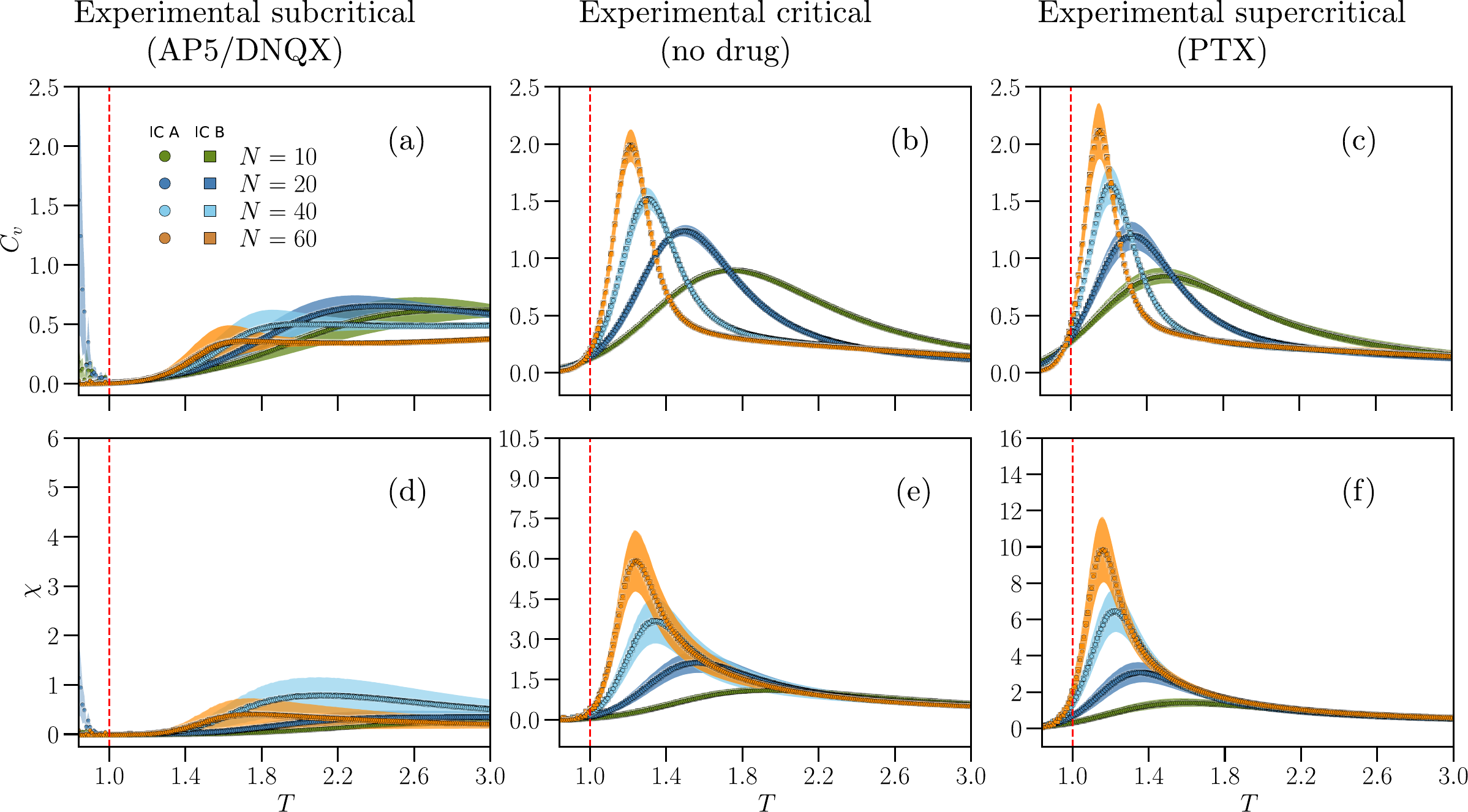}         
    \caption{Same as in Fig.\ (\ref{figS:thermodynamics-IC-comparison}) but for the Ising-like models inferred from the experimental data ($N = 60$ total electrodes; random subsampling at sizes $N \in \{10,20,40\}$). Shaded areas around the curves of $C_{v}$ and $\chi$ correspond to the standard error obtained from $5$ (b, c, e, f) or $4$ (a, d) experimental samples.}         
    \label{figS:thermodynamics-IC-comparison-experimental}        
\end{figure}       

\clearpage

\begin{figure}
    \centering
    \includegraphics[width=0.8\linewidth]{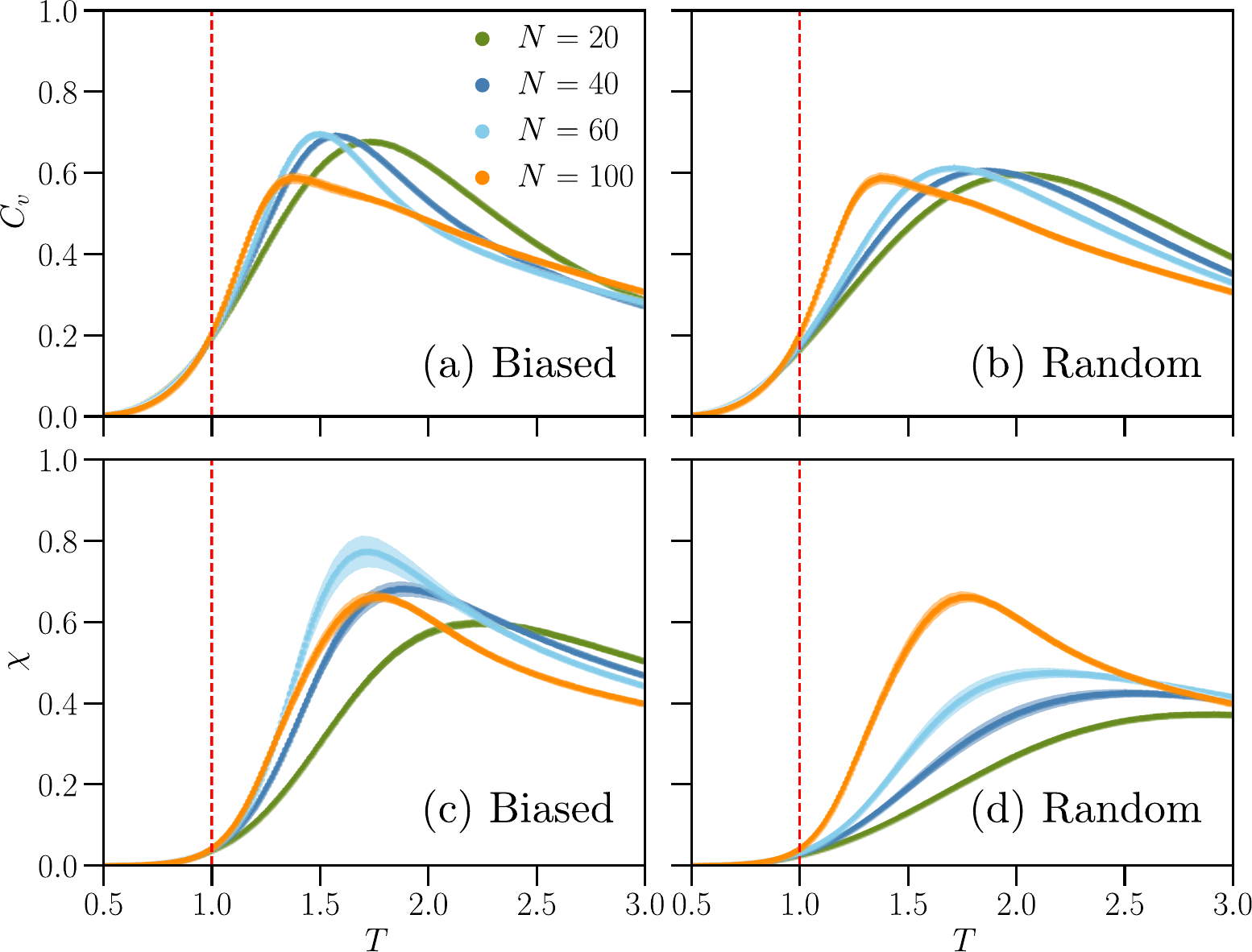} 
    \caption{Specific heat $C_v$ (top row) and susceptibility $\chi$ (bottom row) as functions of temperature $T$ for pairwise Ising models fitted to IF subnetworks $N \in \{20,40,60\}$ of a network with a total of 100 neurons in the subcritical state. Subnetworks were selected either by (a, c) selecting the neurons with the strongest average absolute correlations, $C_i = \frac{1}{N-1} \sum_{j \neq i} |C_{ij}|$, or (b, d) selecting neurons randomly. Since $\PKData$ is not significant for subcritical networks, i.e.\ $V_K \approx 0$ (Fig.\ \ref{fig:parameters-VK}a in the main text), in this case we fit only $\aveSigmaData$ and $\CijData$ for the subnetworks. The data for $N=100$ (orange curves) is the same as in Fig.\ \ref{fig:thermodynamics}a, d of the main text.}    
    \label{figS:thermo-subsampling} 
\end{figure}

\clearpage

\section*{Supplementary tables}

\vspace{2cm} 

\begin{table}[!h]
  \centering
  \caption{Values of the tuning parameter $\durecCrit$ of the IF model that sets the system in the critical state, as a function of the system size $N$.}
  \renewcommand{\arraystretch}{1.3} 
  \setlength{\tabcolsep}{3pt}       
  \begin{tabular}{
    >{\raggedright\arraybackslash}p{0.25\columnwidth} 
    >{\centering\arraybackslash}p{0.25\columnwidth}
    }
    \hline
    \textbf{System size $N$} & \textbf{$\durecCrit$} \\[0.4ex]
    \hline
    20      & $2.50 \cdot 10^{-3}$ \\
    40      & $1.90 \cdot 10^{-3}$ \\
    80      & $1.50 \cdot 10^{-3}$ \\
    100     & $1.26 \cdot 10^{-3}$ \\
    5000    & $1.00 \cdot 10^{-4}$ \\
    8000    & $7.80 \cdot 10^{-5}$ \\
    10000   & $6.70 \cdot 10^{-5}$ \\
    20000   & $4.60 \cdot 10^{-5}$ \\
    30000   & $3.50 \cdot 10^{-5}$ \\
    60000   & $2.30 \cdot 10^{-5}$ \\
    100000  & $1.70 \cdot 10^{-5}$ \\
    \hline
  \end{tabular}
  \label{tableS:durec}
\end{table} 

\vspace{2cm}

\begin{table}[!h]   
  \centering
  \caption{Values of the learning rate parameter $\alpha$ used for the BM learning as a function of the size $N$ and the state of the dynamics of the data. The learning rate for the fields $h_i$ and for the potential $V_K$ is set to $\BMLearningRate(n) = (n+100)^{-\alpha}$. The additive factor of $100$ inside the parenthesis is used to prevent a fast initial learning rate and avoid divergences at the start of the learning procedure. We set a smaller learning rate $\BMLearningRate_J = \BMLearningRate / 2$ for the interaction constants $J_{ij}$ since their number ($ \sim N^2 $) is much larger when compared to the number of fields $h_i$ or the number of potentials $V_K$ ($\sim N$), so we update their values at a slower rate to avoid unfeasibly long CPU times due to instabilities during the learning procedure.}
  \footnotesize
  \renewcommand{\arraystretch}{1.08}
  \setlength{\tabcolsep}{6pt}
  \begin{tabular*}{\textwidth}{@{\extracolsep{\fill}} l c c c c c c @{}}
    \toprule
    \textbf{System size $N$} &
    \shortstack[c]{$\alpha$\\(IF subcritical)} &
    \shortstack[c]{$\alpha$\\(IF critical)} &
    \shortstack[c]{$\alpha$\\(IF supercritical)} &
    \shortstack[c]{$\alpha$\\(Experimental\\subcritical)} &
    \shortstack[c]{$\alpha$\\(Experimental\\critical)} &
    \shortstack[c]{$\alpha$\\(Experimental\\supercritical)} \\
    \midrule 
    10  & --- & --- & --- & 0.5 & 0.7 & 0.9 \\ 
    20  & 0.6 & 0.6 & 0.9 & 0.5 & 0.7 & 0.9 \\
    40  & 0.6 & 0.6 & 0.9 & 0.5 & 0.7 & 0.9 \\
    60  & --- & --- & --- & 0.6 & 0.7 & 0.9 \\
    80  & 0.6 & 0.6 & 1.2 & --- & --- & --- \\
    100 & 0.6 & 0.7 & 1.5 & --- & --- & --- \\
    \bottomrule
  \end{tabular*}
  \label{tableS:learning-rate-parameter}
\end{table}  

\begin{table}[ht]
  \centering
  \caption{Values of $\Tmax$ for which the thermodynamic response functions attain a maximum in the respective curves of Figs.\ \ref{fig:thermodynamics}b,c,e,f (main text), Figs.\ \ref{figS:AP5-results}h,i, and Figs.\ \ref{fig:thermodynamics-experimental}b,c,e,f (main text).} 
  \footnotesize
  \renewcommand{\arraystretch}{1.08}
  \setlength{\tabcolsep}{6pt}
  \begin{tabular*}{\textwidth}{@{\extracolsep{\fill}} l c c c c c @{}}
    \toprule
    \multicolumn{6}{c}{\textbf{Specific heat $C_{v}$}} \\
    \midrule
    \textbf{System size $N$} &
    \shortstack[c]{$\Tmax$\\(IF critical)} &
    \shortstack[c]{$\Tmax$\\(IF supercritical)} &
    \shortstack[c]{$\Tmax$\\(Exp.\ subcritical -- AP5)} & 
    \shortstack[c]{$\Tmax$\\(Exp.\ critical)} &
    \shortstack[c]{$\Tmax$\\(Exp.\ supercritical)} \\
    \midrule
    10  & --- & --- & --- & 1.75 & 1.49 \\ 
    20  & 1.17 & 1.10 & --- & 1.50 & 1.32 \\
    40  & 1.15 & 1.08 & --- & 1.30 & 1.21 \\
    60  & ---  & ---  & 1.26 & 1.21 & 1.15 \\
    80  & 1.11 & 1.04 & --- & --- & --- \\
    100 & 1.11 & 1.05 & --- & --- & --- \\
    \midrule
    \multicolumn{6}{c}{\textbf{Susceptibility $\chi$}} \\
    \midrule
    \textbf{System size $N$} &
    \shortstack[c]{$\Tmax$\\(IF critical)} &
    \shortstack[c]{$\Tmax$\\(IF supercritical)} &
    \shortstack[c]{$\Tmax$\\(Exp.\ subcritical -- AP5)} & 
    \shortstack[c]{$\Tmax$\\(Exp.\ critical)} &
    \shortstack[c]{$\Tmax$\\(Exp.\ supercritical)} \\
    \midrule
    10  & --- & --- & --- & 1.92 & 1.57 \\ 
    20  & 1.32 & 1.05 & --- & 1.56 & 1.35 \\
    40  & 1.23 & 1.09 & --- & 1.35 & 1.22 \\
    60  & ---  & ---  & 1.29 & 1.24 & 1.16 \\
    80  & 1.16 & 1.05 & --- & --- & --- \\
    100 & 1.15 & 1.06 & --- & --- & --- \\
    \bottomrule
  \end{tabular*}
  \label{tableS:Tmax}
\end{table}  

\end{document}